\documentclass[aps,pre,twocolumn,superscriptaddress,10pt,nofootinbib,groupedaddress]{revtex4-1}

\usepackage{subfigure}
\usepackage{amsfonts, amssymb, amsmath}
\usepackage{bm}
\usepackage{graphics}
\usepackage{graphicx}
\usepackage{wasysym}
\usepackage{natbib}
\usepackage{mathtools}

\usepackage[utf8x]{inputenc}
\usepackage[T1]{fontenc}

\usepackage{esint}
\usepackage{bbold}
\usepackage{paralist}
\usepackage{xr}
\usepackage{efbox}

\usepackage[usenames,dvipsnames]{xcolor}

\usepackage{color}
\definecolor{darkblue}{rgb}{0,0,0.6}
\definecolor{darkred}{rgb}{0.6,0,0}

\usepackage{tikz}
\usetikzlibrary{matrix,patterns}
\usetikzlibrary{decorations.pathreplacing}
\usepackage{hyperref}
\hypersetup{
bookmarksopen=true,
pdftitle= Absence of Criticality in the Phase Transitions of Open Floquet Systems,
pdfauthor= Steven Mathey and Sebastian Diehl,
pdftoolbar=false,
pdfstartview={FitH},
pdfmenubar=true,
pdfhighlight=/O,
colorlinks=true,
urlcolor=darkblue,
citecolor=darkblue,
linkcolor=MidnightBlue,
}

\newcommand{\I}{\text{i}}
\newcommand{\ie}{i.e.~}
\newcommand{\Ie}{I.e.~}
\newcommand{\eg}{e.g.~}

\newcommand{\Fig}[1]{Fig.~\ref{#1}}

\newcommand{\Eq}[1]{Eq.~(\ref{#1})}
\newcommand{\eq}[1]{(\ref{#1})}
\newcommand{\Eqs}[1]{Eqs.~(\ref{#1})}
\newcommand{\Sect}[1]{Sect.~\ref{#1}}

\usepackage{abbrevs}
\usepackage{etoolbox}
\newabbrev\RG{Renormalisation Group (RG)}[RG]
\newabbrev\FBZs{Floquet-Brillouin zones (FBZ)\vphantom{\FBZ}}[FBZs]
\newabbrev\FBZ{Floquet-Brillouin zone (FBZ)\vphantom{\FBZs}}[FBZ]
\newabbrev\WF{Wilson-Fisher (WF)}[WF]
\newabbrev\ThreeD{three dimensions (3$d$)}[3$d$]

\robustify{\RG}
\robustify{\FBZ}
\robustify{\FBZs}
\robustify{\WF}
\robustify{\ThreeD}

\makeatletter 
\renewcommand\maybe@space@{%
  \maybe@ictrue 
  \expandafter   \@tfor
    \expandafter \reserved@a
    \expandafter :%
    \expandafter =%
                 \nospacelist
                 \do \t@st@ic
  \ifmaybe@ic 
    \space
  \fi
}
\makeatother

\begin{document}

\title{Absence of Criticality in the Phase Transitions of Open Floquet Systems}

\author{Steven Mathey}
\email[]{smathey@thp.uni-koeln.de}

\author{Sebastian Diehl}

\affiliation{Institut f\"ur Theoretische Physik, Universit\"at zu K\"oln, 50937 Cologne, Germany}

\date{\today}

\begin{abstract}
We address the nature of phase transitions in periodically driven systems coupled to a bath. The latter enables a synchronized non-equilibrium Floquet steady state at finite entropy, which we analyse for rapid drives within a non-equilibrium \RG approach. While the infinitely rapidly driven limit exhibits a second order phase transition, here we reveal that fluctuations turn the transition first order when the driving frequency is finite. This can be traced back to a universal mechanism, which crucially hinges on the competition of degenerate, near critical modes associated to higher Floquet Brillouin zones. The critical exponents of the infinitely rapidly driven system -- including a new, independent one -- can yet be probed experimentally upon smoothly tuning towards that limit.
\end{abstract}

\maketitle

\textit{Introduction} -- Many-body Floquet systems \cite{Eckardt2017,Moessner2017} -- ensembles of particles subject to periodic driving -- have recently triggered enormous research interest, both experimentally and theoretically. For example, very rapid drive can lead to effective conservative dynamics on short enough time scales, as was successfully exploited for Hamiltonian engineering of artificial gauge fields for ultracold atoms \cite{Struck2013,Goldman2014}. When instead the driving frequency $\Omega$ is closer to the natural energy scales of the problem, phenomena directly tied to driving can be observed, such as time crystals in atomic \cite{Choi2017} and ionic \cite{Zhang2017} systems. Theoretical research spans the question of equilibration \cite{DAlessio2014,Lazarides2014,Shirai2015,Chandran2016,Canovi2016,Bukov2015b,Khemani2016,Genske2015,Kuwahara2016,Mori2016,Shirai2016,Weidinger2017,Shirai2018,Howell2018}, the search for novel topological states without equilibrium counterparts \cite{Kitagawa10,Lindner2011,Cayssol2012,Rudner2013,Karzig2015}, or driven analogs of many-body localization \cite{DAlessio2013,Lazarides2015,Ponte2015}.

Specifically when it comes to implementations of periodically driven quantum systems with generic interactions, the ensuing irreversibility can lead to unbounded heating \cite{DAlessio2014,Lazarides2014,Chandran2016,Canovi2016,Bukov2015b,Genske2015,Kuwahara2016,Mori2016,Bukov2015c,Weidinger2017,Clark2017,Kandelaki2017,Howell2018,Shibata2018}. This represents an important hurdle to experimental implementation of many of the anticipated phenomena. A natural cure is to couple the driven system to a bath, such that the system can reach a Floquet steady state, with observables synchronized to the drive. Often such baths occur quite naturally, such as phonons in solid state superfluids \cite{Dehghani2014,Knap2016,Babadi2017,Murakami2017,Seetharam2018}, quantum dots and optical cavities \cite{Xu2015,Chitra2015,Lemonde2016,Stehlik2016,Gong2018}, Brownian motors \cite{Hanggi2009,Salger2013,Denisov2014}, spin chains \cite{Seetharam2015,Lazarides2017,Lerose2018} or cold atoms in optical lattices \cite{Li2016,Iwahori2017,Tomita2017}.

\begin{figure}[ht]
\begin{center}
 \begin{tikzpicture}[scale=.99,transform shape]
\node[] at (0,0) {\includegraphics[width=1\columnwidth]{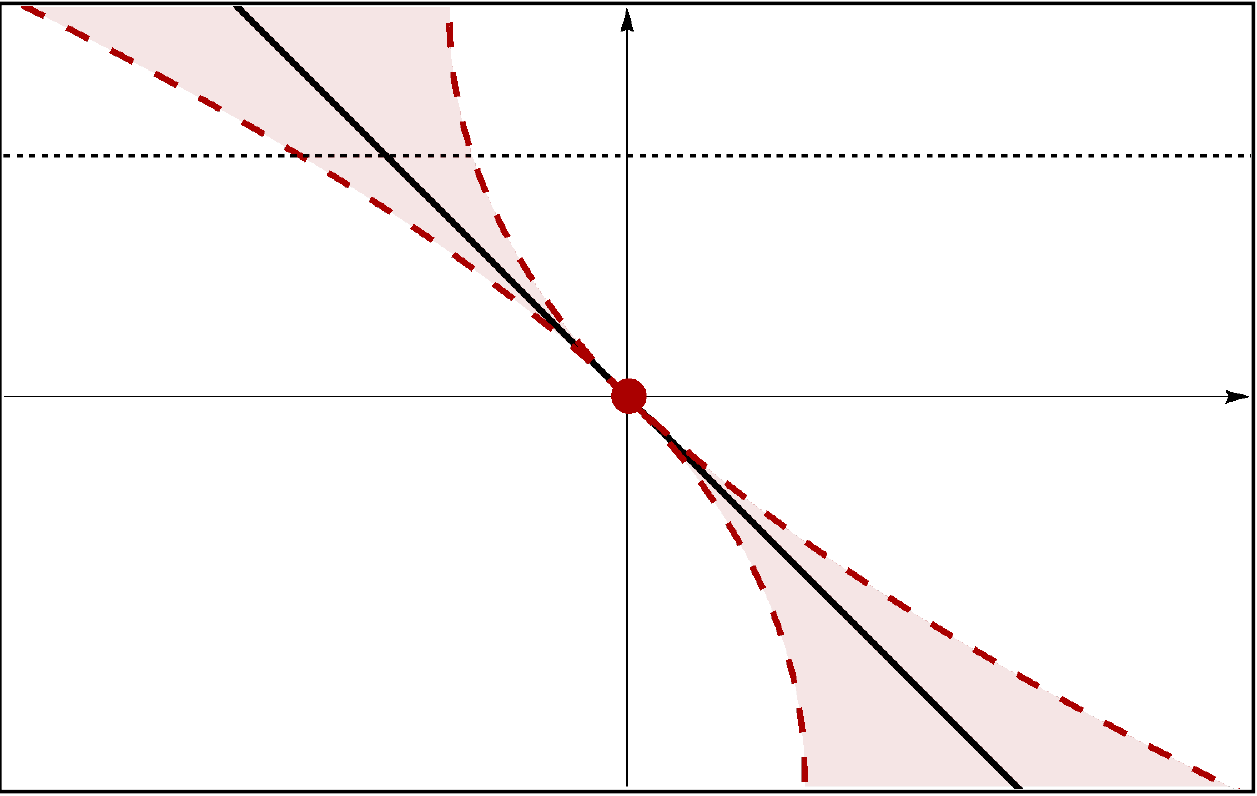}};
\node[] at (-2.6,-1) {Ordered phase};
\node[] at (2.4,1) {Symmetric phase};
\node[] at (4,-0.2) {$\delta t$};
\node[] at (0.5,2.35) {$\hat{x} \sim \frac{1}{\Omega}$};
\draw[<->,thick] (1.2,-1.95) -- (1.87,-1.95);
\node[] at (1.5,-2.15) {$\Delta t$};
\end{tikzpicture}
\end{center}
\caption{Schematic phase diagram of the open Floquet system in $3d$. $\delta t$ is the distance from the phase transition in the infinitely rapidly driven system, $\hat{x} \sim \Omega^{-1}$ is the rescaled drive coefficient  [cf. \Eq{eq_defX}]. The symmetry breaking phase transition occurs at the solid black line. It is second order only at $\Omega^{-1} =0$ (red dot). Otherwise, fluctuations associated to the periodic drive transform the phase transition to weakly first order. The dashed red lines represent a crossover region between the known $\Omega^{-1}=0$ scaling regime and one where scaling is frozen out (light red). The black dotted line represents a typical experimental path through the phase diagram.}
\label{fig_phase_diagram}
\end{figure}
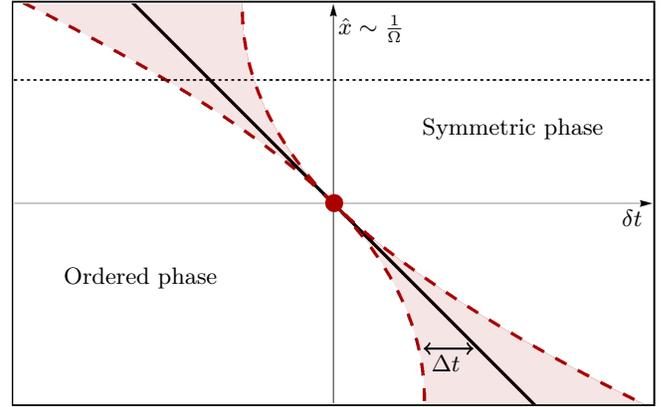

A natural and fundamental question in this large class of periodically driven, open quantum systems concerns the effect of the periodic drive on symmetry breaking phase transitions \cite{DeSarkar2014}. Previous work has addressed this question in the slowly driven limit, establishing the connection to Kibble-Zurek physics \cite{Nikoghosyan2016,Baoquan2016}, as well as intermediate driving frequencies~\cite{Korniss2000,Fujisaka2001,Buendia2008,Lerose2018}. The effect of fast, but not infinitely fast driving remained elusive so far.

In this work, we focus on a minimal model for a rapidly periodically driven open quantum system with phase rotation symmetry in \ThreeD. We identify a universal mechanism, according to which a seeming second order phase transition is unavoidably driven first order by fluctuations.

\emph{Basic physical picture} -- At first sight, the qualitative modification of the critical behavior by a fast scale may appear counterintuitive. It can be rationalized, however, when taking into account the fact that energy is not conserved in open Floquet systems. For any mode with a given frequency, there is a tower of modes with the same frequency but shifted by integer multiples of $\Omega$. This represents the possibility of exchanging energy quanta $n\Omega$ with the driving field  -- a notion of `high' and `low' energies, or `slow' and `fast' modes, is thus not well defined \emph{a priori}.

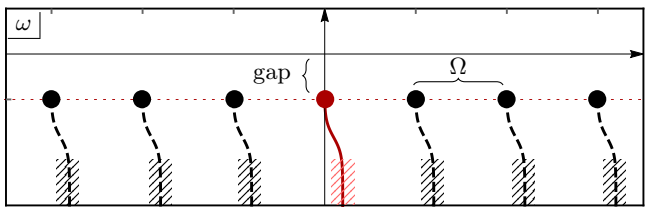
\begin{figure}[t]
\begin{center}
 \begin{tikzpicture}[scale=.99,transform shape]
\node[] at (0,0) {\includegraphics[width=\columnwidth]{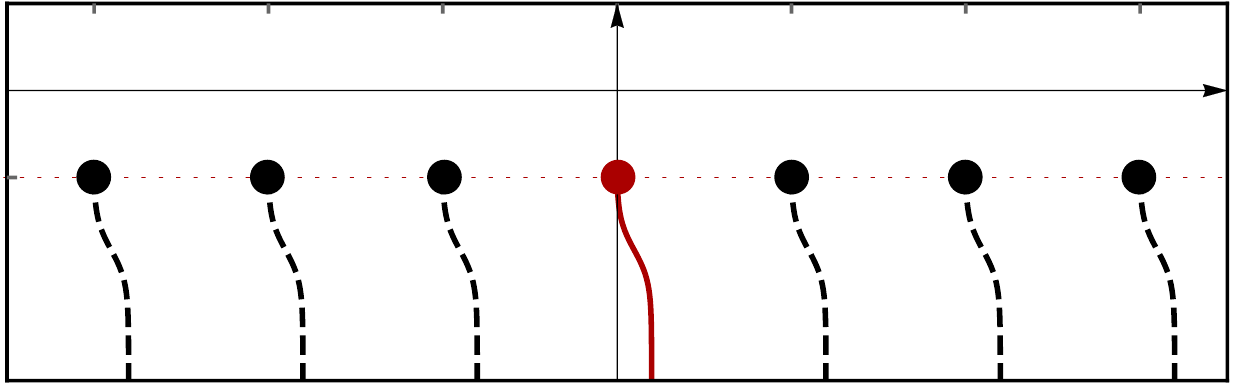}};
\draw [decorate,decoration={brace}] (1.2,0.3) -- (2.4,0.3) node [black,midway,yshift=0.25cm] {$\Omega$};
\draw [decorate,decoration={brace}] (-0.2,0.2) -- (-0.2,0.65) node [black,midway,xshift=-0.5cm] {gap};
\node[] at (-4.03,1.1) {\efbox[leftline=false,topline=false]{$\omega$}};
\fill[pattern=north east lines,pattern color=red!60!White] (0.1,-1.3) rectangle (0.4,-0.7);
\fill[pattern=north east lines,pattern color=black] (1.3,-1.3) rectangle (1.6,-0.7);
\fill[pattern=north east lines,pattern color=black] (2.525,-1.3) rectangle (2.825,-0.7);
\fill[pattern=north east lines,pattern color=black] (3.75,-1.3) rectangle (4.05,-0.7);
\fill[pattern=north east lines,pattern color=black] (-1.15,-1.3) rectangle (-0.85,-0.7);
\fill[pattern=north east lines,pattern color=black] (-2.35,-1.3) rectangle (-2.05,-0.7);
\fill[pattern=north east lines,pattern color=black] (-3.6,-1.3) rectangle (-3.3,-0.7);
\end{tikzpicture}
\end{center}
\caption{Location of the poles of the retarded Green function in the complex frequency plane. The absence of energy conservation gives rise to lines of poles spaced by $\Omega$. The imaginary parts of the pole is the damping rate of the corresponding mode. In a Floquet system, all the modes have the same damping rate and reach criticality simultaneously.}
\label{fig_poles}
\end{figure}

Let us first consider the undriven situation for a general open system.
The proper object to characterize criticality is the retarded single-particle Green function, \Eq{eq_exp}. In the frequency and momentum domain it takes the form
\begin{align*}
G_{R;0}(\omega,\boldsymbol{q}) = \frac{1}{\omega - \epsilon_{\boldsymbol{q}} - \I \gamma_{\boldsymbol{q}}} \, ,
\end{align*}
where we have absorbed the quasiparticle residue in the definition of the energy $\epsilon_{\boldsymbol{q}}$ and damping rate $\gamma_{\boldsymbol{q}}$. In general both $\epsilon_{\boldsymbol{q}}$, but also $\gamma_{\boldsymbol{q}}$ are momentum dependent, continuous functions. Within our model, they are given by ${\epsilon_{\boldsymbol{q}} + \I \gamma_{\boldsymbol{q}} = - K q^2 - \mu_0}$. The poles of $G_{R;0}$ depend on $\boldsymbol{q}$ and thus lie on a line in the complex frequency plane (central solid red line in \Fig{fig_poles}). The imaginary part of the end-point of the line (red dot) represents the system gap -- it provides the decay rate for the slowest mode of the system.
Tuning to criticality is then achieved by making this gap vanish, which happens when the line of poles touches the real axis. A renormalization procedure is then needed to control the singularities induced by the vanishing of the gap, but is well defined in the undriven open system case: It can be designed to gradually integrate out modes with decreasing $q$ along the lines of poles (red overshadowed range).

The situation is drastically different for a periodically driven open system: Now poles are located not only on one central line, but also on all copies of that line shifted by integer multiples of $\Omega$ (dashed lines in \Fig{fig_poles}), according to Floquet's formalism. In particular, when the system becomes critical, all lines extend jointly towards the real axis. Then, the usual strategy of integrating out high energy scales to find the effective low energy theory has to be carefully adapted: The coarse-graining has to take place within each of the lines of poles; but in principle, all the critical poles associated to different lines have to be taken into account. Small scales are therefore integrated out as before, but fast scales remain even at criticality.

We find, however, that the contributions from additional poles are parametrically suppressed for a weak and fast drive. We take advantage of this, and devise an expansion in powers of $\Omega^{-1}$. In addition, we work at one-loop order, which is systematic to first order in powers of $\epsilon = 4-d$ \cite{footnote4}. Our approach is a double expansion, and systematic at $\mathcal{O}(\Omega^{-1}) \times \mathcal{O}(\epsilon)$.

The fact that phase transitions can be driven from second to first order by strong fluctuations occurs also in other contexts. One class is provided by the Coleman-Weinberg or Halperin-Lubensky-Ma mechanism, where additional gapless modes -- such as gauge fields \cite{Coleman1973,Halperin1974} or Goldstone modes \cite{Fisher1974,Nelson1974} -- compete with the critical ones in the vicinity of a phase transition. A second class derives from the Potts model, where the common prerequisite is that a continuous \emph{external} (order parameter) symmetry is explicitly broken down to a non-trivial discrete subgroup (\eg $U(1) \to Z_3$ in the Potts model \cite{Golner1973,Wu1982}, or similar phenomena in $O(N)$ models \cite{Carmona2000,Aharony2003}). This allows for new operators that may turn out to be relevant. Here we reveal another class, where a continuous \emph{internal} symmetry (time translation invariance) is broken down to a discrete one  -- while keeping the external phase rotation symmetry $U(1) \simeq O(2)$ fully intact. Since discrete time translation invariance and energy conservation modulo $\Omega$ are two sides of the same coin, this provides an alternative, RG based viewpoint on our mechanism.

\textit{Open Floquet dynamics} -- Microscopically, our system is made of generic interacting particles on a lattice, governed by a Hamiltonian with a bounded energy spectrum, and coupled to an external bath. The periodic time-dependence typically occurs in the Hamiltonian $H(t+2\pi/\Omega) = H(t)$, but it could also enter through periodic excitations of the bath. The dynamics have a $U(1)$ phase rotation symmetry, also respected by the drive. Our focus will be on phase transitions in \ThreeD systems, where the phase rotation symmetry is broken spontaneously. In the absence of drive, these are continuous, and correspond to critical points where the order parameter has strong large-scale fluctuations that overwhelm the microscopic degrees of freedom. We therefore employ an effective semiclassical, mesoscopic Landau-Ginzburg-type model, where only the dynamics of the complex order parameter $\phi$ is taken into account quantitatively \nocite{Vorberg2013,Vorberg2015,Vorberg2018}\cite{footnote3}. The ensuing stochastic dynamics is governed by the Langevin equation,
\begin{align}
& \I \partial_t \phi = \left[ K \nabla^2  - \mu  - g |\phi|^2 \right] \phi + \xi \, .  \label{eq_ddgpe}
\end{align}
$\xi$ is a Gaussian white noise, which has correlation ${\langle \xi(t,\boldsymbol{x}) \xi^*(t',\boldsymbol{x}') \rangle = 2 \gamma \delta(t-t') \delta(\boldsymbol{x}-\boldsymbol{x}')}$, with $\gamma>0$, and vanishes on average.

The couplings $K,\mu,g$ are complex valued. Their real parts account for the coherent dynamics inherited from the underlying Hamiltonian, and the coupling to the bath is responsible for their imaginary parts \cite{Carusotto2013a,Sieberer2015b}. These determine the phase structure of the system's stationary state. In particular, a second order phase transition accompanied with the spontaneous breakdown of phase rotation symmetry occurs in the undriven system when $\text{Im}(\mu)$ is lowered below its critical value. In our case, all couplings are time-dependent with period $2\pi/\Omega$, as a consequence of the microscopic drive. For definiteness, we choose a monochromatic drive
\begin{align}
& \mu =  \mu_0 + \mu_1 \, \text{e}^{-\I \Omega t} + \mu_{-1} \, \text{e}^{\I \Omega t} \, , \nonumber \\
& g =  g_0 + g_1 \, \text{e}^{-\I \Omega t} + g_{-1} \, \text{e}^{\I \Omega t} \, ,
\label{eq_def_mn_gn}
\end{align}
with $\gamma$ and $K$ constants \cite{footnote5}.

An effectively time-independent, yet driven-dissipative model emerges not only when $\mu_{\pm1}= g_{\pm1} =0$, but also in the limit of infinitely fast driving $\Omega \to \infty$. This limit is appropriate for typical settings in quantum optics, or quantum optical many-body systems \cite{Carusotto2013a,Daley2014,Sieberer2015b}. In that case, the driving scale is approximated as infinitely fast $\Omega^{-1} =0$ (rotating wave approximation). This problem exhibits a true second order phase transition, but a modified criticality compared to equilibrium due to the microscopic breaking of detailed balance \cite{Tauber2013a,Sieberer2013a}.

Here we focus on weakly and rapidly driven Floquet systems, where the driving frequency $\Omega$ is large, but still of comparable order to the other energy scales of the problem. Technically, we incorporate the leading rotating wave corrections $\mathcal{O}(\Omega^{-1})$ into the analysis of the near-critical driven open many-body problem.

\emph{Action and symmetries} -- We re-write the stochastic dynamics of \Eq{eq_ddgpe} in terms of a dynamical functional integral \cite{kamenev2011field,Sieberer2015b}, using the effective action $\Gamma[\Phi]$,
\begin{align}
 \text{e}^{\I \Gamma[\Phi]} = \int {D\varphi} \, \text{e}^{\I \left( S[\Phi+\varphi] + \int_{t,\boldsymbol{x}} \varphi \, \delta \Gamma[\Phi]/\delta \Phi \right)} \, ,
 \label{eq_generating_functional}
\end{align}
which includes all the field fluctuations, and provides the correlation and response functions. \Eq{eq_ddgpe} translates to the mesoscopic action
\begin{align}
 S =  \int_{t,\boldsymbol{x}} \Phi^\dagger \left(\begin{array}{cc} 0 & G_A^{-1} \\ G_R^{-1} & P_K \end{array} \right) \Phi  +  \left( g \, \tilde{\phi}^* \phi \left|\phi\right|^2   + \text{c.c.} \right) \, ,
 \label{eq_keldysh_action}
\end{align}
with $G_R^{-1} = \I \partial_t - K \nabla^2 + \mu$, $G_A^{-1} = \I \partial_t - K^* \nabla^2 + \mu^*$ and $P_K = \I \gamma$. ${\Phi = (\phi,\tilde{\phi})}$ contains the order parameter $\phi$, as well as the 'response' or 'quantum' field $\tilde{\phi}$ that is inherent to the dynamical functional formalism. The following symmetry considerations will guide our understanding:\begin{inparaenum}[(i)]

\item Discrete time translations: Continuous time translations are implemented by ${\Phi (t) \to \Phi (t + \Delta t)}$ for arbitrary $\Delta t$. A drive with frequency $\Omega$, breaks this continuous symmetry down to a discrete one, ${\Delta t = 2\pi n/\Omega}$ with $n$ integer. The continuous symmetry is restored in the undriven limit ${\mu_{\pm 1}= g_{\pm 1}=0}$, but also in the infinitely rapidly driven limit $\Omega^{-1} =0$, where the rotating wave approximation is applicable. Conversely, its explicit breaking allows for the presence of additional dimensionful couplings $\mu_{\pm1}$ and $g_{\pm1}$. These are not compatible with the undriven dynamical $\phi^4$ theory, and will lead to a new relevant direction at the \WF fixed point.

\item Absence of detailed balance: Thermodynamic equilibrium can be formulated in terms of a dynamical symmetry, whose presence is equivalent to the obedience of thermal fluctuation-dissipation relations, \ie detailed balance \cite{tauber2014critical,Sieberer2015,Aron2018}. Out of equilibrium, this symmetry is generically lost. It can, however, formally be recovered by fine-tuning the drive and the dissipation. In our case, this would amount to having the ratios of all pairs of complex couplings to be both real and time independent (see \Sect{app_equilibrium_symmetry}). Whenever this unnatural fine-tuning is not realized, we will encounter the effect described in this work. In this sense, it is generic, or universal, for periodically driven, open quantum systems.
\end{inparaenum}

\emph{Single-particle Green functions and critical poles} -- The Wigner representation \cite{Arrachea2005,wu2008,Stefanucci2008,Tsuji2008,Genske2015} of the single-particle Green functions $G_n(\omega)$, is the double Fourier transform of the real-time Green functions $G(t,t')$ (See \Sect{app_green_definitions}). The discrete time-translation invariance is encoded in the index $n$. The retarded Wigner Green function $G_{R;n}(\omega)$ is composed of an infinite sum of poles located on lines in the complex plane (see \Fig{fig_poles} and \Sect{app_poles}). The residues of the poles of $G_{R;n}(\omega)$ are organised in a power series in $\mu_{\pm1}/\Omega$. This means that a systematic expansion of the loop corrections in powers of $\Omega^{-1}$ is obtained by expanding the Green functions in powers of $\mu_{\pm1}/\Omega$ before the frequency integrations are performed. To order $1$ in $\mu_{\pm1}/\Omega$, we find
\begin{align}
& G_{R;0}(\omega,\boldsymbol{q}) = h^R_0 (\omega,\boldsymbol{q}) , \hspace{0.2cm}  h^R_0(\omega,\boldsymbol{q}) = (\omega+K \boldsymbol{q}^2 +\mu_0)^{-1}, \nonumber \\
& G_{R;n \neq0}(\omega,\boldsymbol{q}) = - \mu_n h^R_0 (\omega- \frac{n \Omega}{2},\boldsymbol{q})h^R_0(\omega + \frac{n \Omega}{2},\boldsymbol{q}).
\label{eq_exp}
\end{align}
$h_0^R(\omega,\boldsymbol{q})$ describes the fundamental pole in the single-particle Green functions. We emphasize that this expansion still captures the correct pole structure and their location, which is fixed by the Floquet formalism. We see that the Green functions involve poles separated by integer multiples of $\Omega$, that all become critical as the gap closes ${\text{Im}(\mu_0)\to 0}$.

\emph{Perturbation theory} -- As anticipated above, care must be taken when renormalizing the problem, due to the absence of a direct meaning of `high' and `low' energies. More practically, this forces us to keep the various poles on equal footing. This imposes a summation over the \FBZ label $n$ in the diagrammatics. The point is illustrated in the one-loop correction to the self-energy at zero frequency and momentum,
\begin{align}
\Delta \mu_0  = 2 \I  \sum_n \int_{\omega,\boldsymbol{q}} g_n G_{K;-n}(\omega,\boldsymbol{q}) \, .
\label{eq_loop_mu}
\end{align}
Using $G_K = - G_R P_K G_A$ and inserting the expansion \Eq{eq_exp}, we can perform the frequency integration and expand it to $\mathcal{O}(\Omega^{-1})$
\begin{align}
& \Delta \mu_0 = \gamma \int_{\boldsymbol{q}}\frac{1}{\left|\text{Im}(K q^2 + \mu_0)\right|} \left(g_0 + \I x \right) \, , \nonumber \\
& x = \frac{\I}{\Omega} \sum_{n\neq0} \frac{g_{-n} \left(\mu_{n}- \mu_{-n}^*\right)}{n} \equiv \sum_{n\neq0} \tilde{g}_n \, .
 \label{eq_defX}
\end{align}
This shows explicitly the appearance of divergences from the $n=0$ term, describing processes exclusively within the zeroth \FBZ, but also from $n\neq 0$, which describe scattering between different \FBZs enabled by the drive. Along the frequency integral of \Eq{eq_loop_mu}, each pole contributes with the same degree of divergence. For a monochromatic drive however, we find ${\text{Res}(\omega_n) \sim (\mu_{\pm 1}/\Omega)^n}$, which leads to a suppression of terms involving higher \FBZs. Thus, all the \FBZs contribute to the critical physics through these divergences, but interactions between different \FBZs are parametrically small in $\Omega^{-1}$.

\emph{\RG analysis} -- Equipped with the understanding of parametrically small but equally divergent contributions from the coupling to higher \FBZs at leading order in $\Omega^{-1}$, we proceed to the resummation of these divergences in an \RG analysis to study their impact on the critical behavior. We first fix the canonical power counting: We transform spatial and temporal coordinates as ${\hat{\boldsymbol{q}}= \boldsymbol{q}/k}$ and ${\hat{\omega}= \omega /[\text{Im}(K)k^2]}$. The couplings are then rescaled as
\begin{align}
 \hat{\mu}_n = k^{-2} \frac{\mu_n}{\text{Im}(K)} \, , && \hat{g}_n = k^{d-4} \frac{\gamma g_n}{4 \text{Im}(K)^2} \, ,
 \label{eq_rescaling}
\end{align}
(with $\tilde{g}_n$ being rescaled as $g_n$). To keep the argument of the oscillatory functions dimensionless, we also rescale ${\hat{\Omega} = \Omega/[\text{Im}(K)k^2]}$.

In order to assess the relevance of these couplings at the interacting \WF fixed point established at  $\Omega^{-1} =0$ \cite{Sieberer2013a}, we include fluctuations into our \RG analysis. To this end, we work at leading order in the $\epsilon = 4-d$ expansion, which requires to include one-loop corrections. The \RG flow equations for $\mu$ and $g$ take the form of a coupled set of differential equations for the dependence of the Fourier modes $\mu_n$ and $g_n$, on the running cut-off scale $k$ (see \Sect{app_rg_flow_equations}). To order $\Omega^{-1}$, the \RG flow equations of $\hat{\mu}_0$ and $\hat{g}_0$ are
\begin{align}
 & k \partial_k \hat{g}_0 = - \epsilon \hat{g}_0 + \frac{10 S_d}{\left|1+\hat{\mu}_0\right|\left(1+\hat{\mu}_0\right)}  \hat{g}_0 \left(\hat{g}_0 +  \sum_m \hat{\tilde{g}}_{m\neq0} \right) \, ,\nonumber \\ 
 & k \partial_k \hat{\mu}_0 = -2 \hat{\mu}_0 - \frac{4S_d}{\left|1+\hat{\mu}_0\right|} \left( \hat{g}_0 +\sum_{m\neq0} \hat{\tilde{g}}_m\right) \, ,
 \label{eq_dimensionless_rg_flow}
\end{align}
with ${S_d = 2 \pi^{d/2}/[(d/2-1)! (2\pi)^{d}]}$. The drive parameter is $\hat{x}=\sum_m \hat{\tilde{g}}_m$. Here and in the following we have simplified our system to make the computation more transparent: We choose $K$, $\mu$ and $g$ to be purely imaginary. Physically, this anticipates the decoherence that occurs in the vicinity of the phase transition, where all coherent dynamics fades away under coarse graining \cite{Sieberer2013a}. We have extracted a factor $\I$ from $\mu_0$, $g_0$ and $K$. The couplings were renamed as ${\mu_0= \I \mu'_0}$, ${g_0 = \I g'_0}$ and ${K = \I K'}$ with $\mu_0'$, $g_0'$ and $K'$ real. We omit the primes to simplify the notation.

In principle, additional variables must be taken into account to compute the \RG flow of $\hat{x}$, since it depends on all the harmonics of $\mu$ and $g$ [see \Eq{eq_defX}]. However, as we show in the \Sect{app_rg_flow_equations_monochromatic}, the loop corrections to the flow of $x$ can be neglected at $\mathcal{O}(\Omega^{-1}) \times \mathcal{O}(\epsilon)$, giving rise to simple dimensional running
\begin{align}
  & k\partial_k \hat{x} = -\epsilon \hat{x} \, .
  \label{eq_x_flow}
\end{align}

The \RG flow equations \eq{eq_dimensionless_rg_flow} and \eq{eq_x_flow} provide a generalisation of the well known, time translation invariant, \RG flow. Indeed, the \WF fixed point emerges when $\hat{x}=0$ (and $\hat{\mu}_0=\mu^*$ and $\hat{g}_0=g^*$). Our analysis reveals that the periodic drive gives rise to a new relevant coupling. In the absence of continuous time translation invariance, the critical point is thus bicritical: Two fine-tunings are necessary to reach it, and to reveal its critical scaling properties. Thus, when tuning across the symmetry breaking phase transition at finite $\Omega^{-1}$ (along the dotted line of \Fig{fig_phase_diagram}) the additional relevant direction provides a finite correlation length. Moreover, in the absence of drive and far away from the critical point, the system is either in a disordered or an ordered phase. This property is robust for a finite, rapid drive since the Green functions are gapped in these phases (cf. \Fig{fig_poles}), and perturbation theory converges \cite{footnote}. This gives rise to a symmetry breaking phase transition without asymptotic criticality, which must be interpreted as a fluctuation induced first order transition.

The linear stability analysis of the \RG flow equations close to the \WF fixed point provides three quantitative predictions:\begin{inparaenum}[(i)]

\item New scaling exponent: We find three critical exponents: $-2 + 2 \epsilon /5 = -1/\nu$, $\epsilon$ and a new independent exponent $-\epsilon=-1/\nu_d$. The first two are known from the equilibrium system, with the first being negative and corresponding to the relevant direction. When the system is infinitely rapidly driven, it is tuned to criticality by tuning $\mu$ and/or $g$ such that $\delta t = A (\delta g+4 \pi^2 \delta \mu)$ vanishes (with $A>0$ a non-universal constant, $\delta \mu = \hat{\mu}_0-\mu_0^*$ and $\delta g = \hat{g}_0-g_0^*$). Then the correlation length diverges as $\xi \sim \delta t^{-\nu}$. In the presence of a drive however ($\hat{x}\neq0$), the correlation length never diverges. $\delta t$ can be tuned to maximize it (or, in \RG terms, bring the flow as close as possible to the \WF fixed point), but $\xi$ ultimately crosses over to a finite value that scales as $\xi \sim \hat{x}^{-\nu_d}$.

\item Shift of the phase transition: The location of the phase transition is shifted in a non-universal although drive-dependant way. The macroscopic phase is ultimately determined by the sign of ${\Delta t = \delta t + A \hat{x}}$. See \Fig{fig_phase_diagram} and \Sect{app_fixed_point} for additional details.

\item Observability of scaling: The above scaling analysis can be refined by replacing $\delta t$ by $\Delta t$. $\hat{x}$ and $\Delta t$ control the crossover between the two scaling regimes. For $\left|\Delta t\right| \gg \left|\hat{x}\right|^{\nu_d/\nu}$, the undriven relevant coupling dominates and the correlation length scales as $\xi \sim \Delta t^{-\nu}$. When $\left|\Delta t\right| \ll \left|\hat{x}\right|^{\nu_d/\nu}$ the correlation length saturates to $\xi \sim \hat{x}^{-\nu_d}$. This crossover is represented as red dashed lines in \Fig{fig_phase_diagram}. The correlation length scales with $\Delta t$ outside of the light red area and it saturates as the dashed red lines are crossed. In particular, this implies that the new critical exponent $\nu_d$ can be observed by varying $\Omega$.
\end{inparaenum}

\textit{Conclusion} -- There is an interesting 'duality' of our scenario to the paradigmatic Kibble-Zurek phenomenology \cite{kibble1976,Zurek1985}. Both the equilibrium limit of an undriven system $\Omega =0$, and the infinitely rapidly driven limit $\Omega^{-1} =0$, afford time-independent descriptions, and exhibit symmetry breaking continuous phase transitions. Here we have shown that asymptotic scaling is cut off at any finite $\Omega^{-1}$.  The Kibble-Zurek phenomenon occurs in the opposite limit of a slow driving: The non-equilibrium conditions are encoded in a slow quench of the couplings. Then the quench rate is analogous to $x$; it stops the correlation length from diverging. Although the underlying mechanisms are very different, in both cases the critical physics is masked and observable only upon smoothly approaching the extreme limiting cases. We reserve the exploration of this connection to future work.

Another intriguing direction of research concerns the applicability of our results to possible phase transitions in long-lived transient states of Floquet systems not coupled to external baths \cite{Chandran2016,Canovi2016,Bukov2015b,Kuwahara2016,Mori2016,Weidinger2017,Howell2018}.

\textit{Acknowledgments - } We thank A. Altland, C.-E. Bardyn, M. Buchhold, C. Duclut, A. Gambassi, M. Heyl, A. Lazarides, G. Loza, J. Marino, R. Moessner, F. Piazza, A. Polkovnikov, G. Refael, A. Rosch, D. Roscher, M. Scherer, K. Seetharam, U. T\"auber and J. Wilson for useful and inspiring discussions. We acknowledge support by the Institutional Strategy of the University of Cologne within the German Excellence Initiative (ZUK 81), by the funding from the European Research Council (ERC) under the Horizon 2020 research and innovation program, Grant Agreement No. 647434 (DOQS), and by the DFG Collaborative Research Center (CRC) 1238 Project No. 277146847 - project C04.  

\appendix

\section{Green functions}
\label{app_green_functions}

In this section we discuss the Green functions in the Floquet steady state. In \Sect{app_green_definitions} we define the Wigner and Floquet Green functions and show how they are computed in general. In \Sect{app_poles} we compute analytically the single-particle retarded Green function and elucidate its pole structure.

\subsection{Definitions}
\label{app_green_definitions}

We exploit here the fact that physical observables are periodic in their center of mass time in the synchronized Floquet steady state. The connected Green functions
\begin{align}
 G(t,t') = -\I \left(\begin{array}{cc} \langle \phi(t) \phi^*(t') \rangle & \langle \phi(t) \tilde{\phi}^*(t') \rangle \\
                      \langle \tilde{\phi}(t) \phi^*(t')\rangle & \langle \tilde{\phi}(t) \tilde{\phi}^*(t') \rangle
                     \end{array} \right) \, ,
\label{eq_defg}
\end{align}
take the following form,
\begin{align*}
 G(t,t') = \sum_n \int_{\omega,\boldsymbol{q}} \text{e}^{-\I\left[\omega (t-t') - \boldsymbol{q} \cdot \boldsymbol{r} + n \Omega (t+t')/2\right]} \, G_n(\omega,\boldsymbol{q}) \, .
\end{align*}
$\boldsymbol{r} = \boldsymbol{x}-\boldsymbol{x}'$ denotes the relative spatial coordinate, which we do not write explicitly on the left-hand-side. $G(t,t')$ is a periodic function of $t_a=(t+t')/2$ and can be represented in terms of the Wigner Green functions \cite{Arrachea2005,wu2008,Stefanucci2008,Tsuji2008,Genske2015},
\begin{align}
 G_n(\omega,\boldsymbol{q}) = \int_{\tau,\boldsymbol{r}} \fint_{t_a} \text{e}^{\I\left[\omega \tau - \boldsymbol{q} \cdot \boldsymbol{r} + n \Omega t_a \right]} \, G(t_a+\tau/2,t_a-\tau/2) ,
 \label{eq_wigner_greens}
\end{align}
which encode the time periodicity with a discrete index. Here and in the following, we use the short-hand notation ${\int_{\omega,\boldsymbol{q}} = 1/(2\pi)^{d+1} \int_{-\infty}^{\infty}\text{d}\omega \int_{-\infty}^{\infty}\text{d}^dq}$, ${\int_{\tau,\boldsymbol{r}} = \int_{-\infty}^{\infty}\text{d}\tau \int_{-\infty}^{\infty}\text{d}^dr}$ and ${\fint_{t_a} = \Omega/(2\pi) \int_0^{2\pi/\Omega}\text{d}t_a}$. Furthermore, the Floquet Green functions are defined as,
\begin{align}
 G_{nm}(\omega) = G_{n-m}\left(\omega+\frac{(n+m)\Omega}{2}\right) \, , && \left|\omega\right| \leq \frac{\Omega}{2} \, ,
 \label{eq_gnm}
\end{align}
which are two-index Green functions constructed from the single index Wigner Green functions.

These definitions are directly applicable to the inverse Green functions, $G^{-1}$. Specifically, $G^{-1}(t,t')$ is [see \Eq{eq_keldysh_action}]
\begin{align*}
 G^{-1}(t,t') \hspace{-2pt} = \hspace{-2pt} \left( \hspace{-3pt} \begin{array}{cc} 0 & \I \partial_t - K^* \nabla^2 + \mu^* \\
           \I \partial_t - K \nabla^2 + \mu & \I \gamma
          \end{array} \hspace{-3pt} \right) \hspace{-2pt}  \delta(t-t') ,
\end{align*}
and $G_{nm}^{-1}(\omega)$ reads
\begin{align}
  G_{nm}^{-1}(\omega) = & \, \left( \begin{array}{cc} 0 & G_{A;nm}^{-1}(\omega) \\
           G_{R;nm}^{-1}(\omega) & P_{K;nm}(\omega)
           \end{array}\right) \, ,
\label{eq_green_floquet}
\end{align}
with
\begin{align}
G_{R;nm}^{-1} & =  \delta_{nm} \left(\omega + n \Omega \right) + K_{n-m}p^2 + {\mu}_{n-m}  \nonumber \\
& =  \delta_{nm} \left(\omega + n \Omega + M_0 \right) + \delta_{n m+1}{\mu}_{1} + \delta_{n m-1}{\mu}_{-1} \, , \nonumber \\
 G_{A;nm}^{-1} & =  \delta_{nm} \left(\omega + n \Omega \right)+ K^*_{n-m}p^2 + ({\mu}_{m-n})^* \nonumber \\
& = \delta_{nm} \left(\omega + n \Omega + M_0^* \right) + \delta_{n m+1}{\mu}_{-1}^* + \delta_{n m-1}{\mu}_{1}^* \, ,\nonumber \\
 P_{K;nm} & =   \I \gamma_{n-m}   = \I \delta_{nm} \gamma \, .
\end{align}
${\mu}_n$ is the $n^\text{th}$ Fourier mode of $\mu$. The expressions after the second equalities are specific to a monochromatic drive with frequency $\Omega$ and $K$ and $\gamma$ constant, and $M_0 = K p^2 + \mu_0$.

The Floquet Green functions are introduced because they provide an efficient means to compute the Green functions from $G^{-1}$. Indeed, $G$ is computed from the inverse Green functions through
\begin{align}
G(t,t') = \left(\begin{array}{cc} - \left[G_R P_K G_A\right](t,t') & G_R(t,t') \\ G_A(t,t') & 0 \end{array} \right) \, ,
\label{eq_greens}
\end{align}
and $G_X$ (with $X=R$ or $A$) is the functional inverse of $G_X^{-1}$, ${\int_\tau G_X^{-1}(t,\tau) G_X(\tau,t') = \delta(t-t')}$. The products that appears in the top-left entry of \Eq{eq_greens} are also functional, ${AB(t,t') = \int_\tau A(t,\tau)B(\tau,t')}$. The Floquet representation of the Green functions \eq{eq_gnm}, has the advantage that it turns functional inverses into matrix inverses. In other words, the following statements are equivalent,
\begin{align}
 & \int_\tau G(t,\tau) G^{-1}(\tau,t') = \delta(t-t') \, , \nonumber \\
 & \sum_{s} G_{ns}(\omega) G^{-1}_{sm}(\omega) = \delta_{nm} \, .
\end{align}
This provides a practical means to compute the Wigner Green functions from $G^{-1}(t,t')$. In particular, we use it later on to expand $G_n(\omega,\boldsymbol{q})$ in powers of $\mu_{n\neq 0 }/\Omega$.

\subsection{Poles}
\label{app_poles}

In this section we compute $G_{R;n}(\omega,\boldsymbol{q})$ analytically and show that it has an infinite number of poles with identical imaginary parts and real parts separated by integer multiples of $\Omega$. The main result of this section is \Eq{eq_sum_with_half_poles} with the additional constraint that, in the sum, only terms where $m$ has the same parity as $n$ contribute. For definiteness, we work with a monochromatic drive, \Eq{eq_def_mn_gn}.

We start by computing the single-particle retarded Green function in real time, $G_R(t,t') = -\I \langle \phi(t) \tilde{\phi}^{*}(t') \rangle$. Without interaction, the solution of \Eq{eq_ddgpe} is
\begin{align}
 \phi(t) = \phi(t_0) \text{e}^{\I \int_{t_0}^t M(t') \text{d}t'} \hspace{-1pt} - \I \int_{t_0}^t \text{e}^{\I \int_{t'}^t M(t'') \text{d}t''}\xi(t') \text{d}t' ,
 \label{eq_real_time_solution}
\end{align}
with $M(t) = Kp^2+\mu(t)$ the real time representation of the right-hand-side of the single-particle equation of motion ($M_0 = Kp^2+\mu_0$ and $M_{n\neq0}=\mu_{n})$. We now exploit the following relation
\begin{align}
 G_R(t,t') = \left. \frac{\delta \langle \phi(t) \rangle_f}{\delta f(t')} \right|_{f=0} \, ,
\end{align}
where $\langle \dots \rangle_f$ is the average over a modified noise ${\xi'=\xi+f}$. This noise has the same Gaussian statistics and variance as $\xi$, but it does not average to zero. Instead we have ${\langle \xi'(t) \rangle_f = f(t)}$. Then, the average of \Eq{eq_real_time_solution} is
\begin{align*}
 \langle \phi(t) \rangle_f = \langle \phi(t_0) \rangle_f \text{e}^{\I \int_{t_0}^t M(t') \text{d}t'} \hspace{-0.1cm}- \I \int_{t_0}^t \text{e}^{\I \int_{t'}^t M(t'') \text{d}t''}  \hspace{-0.1cm} f(t') \text{d}t' ,
\end{align*}
and we obtain
\begin{align}
 G_R(t,t') = -\I \theta(t-t') \text{e}^{\I \int_{t'}^t M(t'') \text{d}t''} \, .
\end{align}
In the Floquet steady state, $t_0 \rightarrow - \infty$ and the term containing $t_0$ is negligible because ${\text{Im}(M_0) >}0$.

We now partially convert $G_R(t,t')$ to its Wigner form
\begin{align}
 G_R(\omega,t_a) = \int_\tau \text{e}^{\I \omega \tau} G_R(t_a+\tau/2,t_a-\tau/2) \, .
\end{align}
To this end, we write
\begin{align*}
 & G_R(t_a+\tau/2,t_a-\tau/2) = \nonumber \\
 & -\I \theta(\tau) \text{e}^{\I M_0 \tau} \text{Exp}\left[\frac{2 \I \sin\left(\Omega \frac{\tau}{2}\right)}{\Omega} \left( M_1 \text{e}^{-\I \Omega t_a} + M_{-1} \text{e}^{\I \Omega t_a}\right) \right] ,
\end{align*}
and use ${\text{e}^{\I z \sin(\theta)} = \sum_m J_m(z) \text{e}^{\I m \theta}}$ (Jacobi-Anger expansion), with $J_m(x)$ the $m^\text{th}$ Bessel function of the first kind,
\begin{align*}
 & G_R(t_a+\tau/2,t_a-\tau/2) = \nonumber \\
 & -\I \theta(\tau) \text{e}^{\I M_0 \tau} \sum_m J_m\bigg[\frac{2\left( M_1 \text{e}^{-\I \Omega t_a} + M_{-1} \text{e}^{\I \Omega t_a}\right)}{\Omega} \bigg] \text{e}^{\I m \Omega \frac{\tau}{2}} .
\end{align*}
Finally we obtain
\begin{align}
 G_R(\omega,t_a) = \sum_m \frac{J_m\bigg[\frac{2\left( M_1 \text{e}^{-\I \Omega t_a} + M_{-1} \text{e}^{\I \Omega t_a}\right)}{\Omega} \bigg]}{\omega+M_0+\frac{m \Omega}{2}}  \, .
 \label{eq_green_omega_t}
\end{align}
From the above equation, it appears that the poles are spaced by half-integer multiples of $\Omega$. This is however not the case because only half of the terms of the above sum contribute to the Wigner Green function ${G_{R;n}(\omega) = \fint_{t_a} \text{e}^{\I n \Omega t_a} G_R(\omega,t_a)}$. To see this, we expand the Bessel function and look at the Fourier transform of each term,
 \begin{align}
 & G_{R;n}(\omega) = \sum_{s=0}^\infty \frac{(-1)^s \Omega^{-2s}}{s!} \Bigg[ \frac{H_{n,0,s}}{s!(\omega+M_0)}  \nonumber \\
 & \hspace{-5pt} + \hspace{-4pt} \sum_{m=1}^{\infty} \frac{\Omega^{-m}}{(s+m)!} \hspace{-2pt} \left( \hspace{-2pt} \frac{ H_{n,m,s}}{\omega+M_0+\frac{m \Omega}{2}} + \frac{(-1)^m  H_{n,m,s}}{\omega+M_0-\frac{m \Omega}{2}} \hspace{-2pt} \right) \hspace{-3pt} \Bigg] ,
 \label{eq_sum_with_half_poles}
\end{align}
 with
\begin{align*}
 H_{n,m,s} & = \fint_{t_a} \text{e}^{\I n \Omega t_a} \left( M_1 \text{e}^{-\I \Omega t_a} + M_{-1} \text{e}^{\I \Omega t_a}\right)^{2s+m} \nonumber \\
 & = \sum_{r=0}^{2s+m} \left(\begin{array}{c} 2s+m \\ r \end{array}\right) M_1^{2s+m-r} M_{-1}^{r} \delta_{n-2s+2r -m} \, .
\end{align*}
The Fourier transform produces a constraint that can only be satisfied if $n+m$ is even. Then $n$ and $m$ must have the same parity and half of the terms in the sum of \Eq{eq_sum_with_half_poles} vanish.

\Eq{eq_sum_with_half_poles} provides an exact expression for the single-particle Green function. Although it is not really useful for practical calculations, it elucidates the analytical structure of the Wigner Green functions. This is most clearly seen in \Eq{eq_green_omega_t}: The frequency dependence of the single-particle Green function is composed of an infinite sum of poles that all have the same imaginary parts and are spaced by integer multiples of $\Omega$. Moreover, the residues of each pole are analytical functions of $M_{\pm1}/\Omega$. Then, expanding $G_{n;R}$ to order $1$ in $M_{\pm1}/\Omega$ provides
\begin{align}
 G_{R;n}(\omega) = \frac{\delta_{n0}}{\omega+M_0} - \frac{M_n (1-\delta_{n0})}{\left(\omega+M_0\right)^2-\left(n\frac{\Omega}{2}\right)^2} \, ,
 \label{eq_exp_app}
\end{align}
which, recalling that $M_0 = Kq^2+\mu_0$ and $M_{n \neq 0} = \mu_{n\neq 0}$, is \Eq{eq_exp}.

\section{RG flow equations}
\label{app_rg_flow_equations}

In this section we show how to obtain \RG flow equations including loop corrections for all the Fourier modes of $\mu$ and $g$. Although we expand the \RG flow equations to order $\Omega^{-1}$, this approach can be applied to include higher order corrections. We recover \Eqs{eq_dimensionless_rg_flow} and \eq{eq_x_flow}.

The \RG flow equations of $\mu$ and $g$ are similar to the undriven case although with additional sums over the Fourier indexes and propagators that are modified by the periodicity of $\mu$. To $1$-loop order (and for arbitrary $\Omega$), the \RG flow equations of $\mu_n$ and $g_n$ take the form
\begin{align}
 k&\partial_k {g}_n = 2 \I S_d k^d \sum_{\substack{m_1,m_2\\m_3,m_4}}  {g}_{m_4} \delta_{n,m_{1234}} \int_\omega \nonumber \\
&  \times \Bigg\{ G_{R;m_1}(\omega,k) G_{K;m_2}\left[-\omega - \frac{m_3-m_4}{2} \Omega ,k\right] {g}_{m_3} \nonumber \\
& \quad + 2 G_{A;m_1}(\omega,k) \, G_{K;m_2}\left[\omega - \frac{m_4-m_3}{2} \Omega ,k\right] ({g}_{-m_3})^* \nonumber \\
 &  \quad + 2 G_{R;m_1}(\omega,k) G_{K;m_2}\left[\omega - \frac{m_3-m_4}{2} \Omega ,k\right] {g}_{m_3}  \Bigg\}\, , \nonumber \\
 k&  \partial_k {\mu}_n  = - 2 \I S_d k^d \sum_m \int_\omega {g}_m G_{K;n-m}(\omega,k) \, ;
 \label{eq_dimensionful_flow_coupling}
\end{align} 
${m_{1234} = \sum_{i=1}^4 m_i}$ and ${S_d = 2 \pi^{d/2}/[(d/2-1)! (2\pi)^{d}]}$. Both equations are obtained by integrating out $1$-loop fluctuations within a momentum shell ${k-dk<p<k}$ and taking the limit ${dk\rightarrow 0}$. $G_{K;n}$, $G_{R;n}$ and $G_{A;n}$ are the Wigner Green functions, which are defined in \Eq{eq_wigner_greens}, and ${g}_n$ are the Fourier modes of $g$. The above equations do not contain any approximation in the Floquet sector. The complexity of the Floquet formalism is hidden in the Wigner Green functions.

We emphasize that the full frequency integral is unavoidable here. The sharp momentum cut-off (inherent to momentum-shell renormalization) fixes the loop momentum to $p=k$ and the frequency integrals are performed by closing the integral path in the complex plane and using the residue theorem. As we discuss in the main text, the poles of the Green functions are located along a line in the complex plane with identical imaginary parts and real parts separated by integer multiples of the drive frequency, ${\omega_n = -K k^2 - \mu_0 + n\Omega}$. See \Sect{app_poles}. In usual \RG approaches, there is only one pole (with $n=0$). Then, the frequency axis is effectively cut-off since only frequencies with $\omega \sim k^2$ contribute to the loop integrals. Only small energies contribute. Although, this is technically very similar to our problem (find the poles and use residue theorem), it is physically very different. Indeed, Floquet formalism forces us to consider all the poles on equal footing and there is no cut-off on the frequency axis.

We now discuss how the \RG flow equations \Eq{eq_dimensionful_flow_coupling}, can be simplified in the presence of a weak and fast drive. This expansion is carried out in three steps: \begin{inparaenum}[(i)]
\item \label{point_mu_expansion} The Wigner Green functions are expanded in powers of $\mu_{n\neq0}$ up to a given order. See \eg \Eq{eq_exp_app}, where $G_{R;n}(\omega)$ is expanded to order $1$. This provides a simplified set of Green functions that are inserted in the \RG flow equations [\Eq{eq_dimensionful_flow_coupling}].
\item The frequency integrals are performed analytically with the residue theorem.
\item The obtained expressions are expanded in powers of $\Omega^{-1}$ up to the same order as the $\mu_{n\neq 0}$ expansion of step (\ref{point_mu_expansion}).
\end{inparaenum}

To order $\Omega^{-1}$, the \RG flow equations are
\begin{align}
& k \partial_k \mu_n = -\frac{S_d k^d \gamma}{\left|\text{Im}(M_0)\right|} \left(g_n + \I X_n \right) \, , \nonumber \\
&  k \partial_k g_{2r+1} = \frac{S_d k^d \gamma}{\left|\text{Im}(M_0)\right|} Y_{2r+1} \, , \nonumber \\
& k \partial_k g_{2r} = \frac{S_d k^d \gamma}{\left|\text{Im}(M_0)\right|} \nonumber \\
& \quad \times \left\{\left[\frac{g_r}{2 M_0} +  \frac{g_r-(g_{-r})^*}{\I  \text{Im}(M_0)} \right] \left(g_r+\I X_r\right) + Y_{2r}\right\} \, .
\label{eq_dimensionful_flow_equations}
\end{align}
Here and in the following we use $M_0 = K k^2 + \mu_0$. $Y_n$ and $X_n$ provide $\mathcal{O}(\Omega^{-1})$ corrections
\begin{align}
 & X_n = \frac{\I}{\Omega} \sum_{m\neq0} \frac{g_{n-m} \left( \mu_{m}-\mu_{-m}^*\right)}{m} \, , \nonumber \\
 & Y_n =  \frac{4}{\Omega}\sum_{m\neq -n/2} \frac{g_{n+m} (g_m)^* }{2m+n} \, .
 \label{eq_defXY}
\end{align}

The calculation of \Sect{app_poles} can be used to argue that the above expansion is systematic in $\Omega^{-1}$: When the \RG flow equations are truncated to a given order in ${\mu_{n\neq 0}}$, then the obtained equations contain all the terms of the same order (and smaller) in $\Omega^{-1}$. This is most readily seen from \Eq{eq_green_omega_t}, where $\mu_{n\neq 0}$ only appears in the argument of the Bessel functions, and is divided by $\Omega$. Expanding this equation in powers of $\mu_{n\neq0}$ is actually an expansion in powers of $\mu_{n\neq0}/\Omega$. Additionally, $\Omega$ appears in the poles of $G_{n;R}(\omega)$. Since these exclusively come with negative powers of $\Omega$, they can not lower the order in $\Omega^{-1}$. When the \RG flow equations are truncated at a given order in $\mu_{n\neq0}/\Omega$, then the poles only produce sub-leading (in $\Omega^{-1}$) contributions. Both types of terms are visible in \Eqs{eq_dimensionful_flow_equations}: $X_n\sim\mu/\Omega$ comes from the argument of a Bessel function and $Y_n \sim g^2/\Omega$ is the $\mathcal{O}(\Omega^{-1})$ (sub-leading) correction of a term of order zero in $\mu_{n\neq0}/\Omega$.

In summary, the above equations are obtained through a double perturbative expansion. \Eqs{eq_dimensionful_flow_coupling} are controlled for a weak coupling. They are systematic to order one in the $\epsilon = 4-d$ expansion, as in a standard $\phi^4$ analysis. \Eqs{eq_dimensionful_flow_equations} are the result of a further expansion in powers of $\Omega^{-1}$ and are systematic to order $1$ as well. Our result are therefore systematic to $\mathcal{O}(\epsilon) \times \mathcal{O}(\Omega^{-1})$.

\subsection{Monochromatic drive}
\label{app_rg_flow_equations_monochromatic}

Here we show that within our mesoscopic model,
\begin{align}
& \mu =  \mu_0 + \mu_1 \, \text{e}^{-\I \Omega t} + \mu_{-1} \, \text{e}^{\I \Omega t} \, , \nonumber \\
& g =  g_0 + g_1 \, \text{e}^{-\I \Omega t} + g_{-1} \, \text{e}^{\I \Omega t}  \, ,
\end{align}
the \RG flow equations can be greatly simplified because the drive is monochromatic. Specifically, we can focus on the $n=0$ sector
\begin{align}
& k \partial_k \mu_0 = -\frac{S_d k^d \gamma}{\left|\text{Im}(M_0)\right|} \left(g_0 + \I X_0 \right) \, , \nonumber \\
& k \partial_k g_{0} = \frac{S_d k^d \gamma}{\left|\text{Im}(M_0)\right|} \nonumber \\
& \quad \times \left\{ \left[\frac{g_0}{2 M_0} +  \frac{g_0-(g_{0})^*}{\I  \text{Im}(M_0)} \right] \left(g_0 + \I X_0\right) + Y_{0} \right\} \, ,
\label{eq_dimensionful_flow_equations_0}
\end{align}
which will lead to a closed set of equations. We remark in passing that the first equation can be derived from \Eq{eq_defX} with $X_0 = x$. To order $\Omega^{-1}$, the flow of $X_0$ and $Y_0$ is given by
\begin{align}
 & k \partial_k X_0 = -\frac{\I S_d k^d \gamma}{2\left|\text{Im}(M_0)\right|} \Bigg\{ Y_0  \nonumber \\
 & + \frac{1}{\Omega} \sum_{m\neq 0} \frac{\left(\mu_{-2m}-\mu_{2m}^*\right)g_m}{m} \left[\frac{g_m}{2 M_0} +  \frac{g_m-(g_{-m})^*}{\I  \text{Im}(M_0)} \right] \Bigg\} \, , \nonumber \\
 & k \partial_k Y_0 = \frac{1}{\Omega} \frac{S_d k^d \gamma}{\left|\text{Im}(M_0)\right|} \nonumber \\
 & \qquad \times \sum_{m\neq0} \frac{g_{2m}^* g_m}{m}\left[\frac{g_m}{2M_0}+\frac{g_m-g_{-m}^*}{\I\text{Im}(M_0)} \right] + \text{c.c.}\, .
 \label{eq_dimensionful_flow_XY}
\end{align}

In the case of a monochromatic drive the last terms in \Eq{eq_dimensionful_flow_XY} can be neglected (then ${k\partial_k Y_0 = 0}$ and ${k\partial_k X_0 \sim Y_0}$) because they contain Fourier modes with $|n|\geq 2$. Indeed, the solution of the \RG flow equations take the form ${\mu_n(k) = \mu_n(\Lambda) + \int_k^\Lambda I(k') \text{d}k'}$ (and similarly for $g_n$). $I(k)$ represents the loop contributions. When $\mu_n(\Lambda)=0$ (as is the case for $|n|\geq2$ and a monochromatic drive), then $\mu_n(k)$ is proportional to $I(k')$. The magnitude of the flow of $X_0$ and $Y_0$ is estimated by estimating $g_{2m}$ to $\mathcal{O}(\Omega^{0})$ and inserting it on the right-hand-side of \Eq{eq_dimensionful_flow_XY}. \Eq{eq_dimensionful_flow_equations} provides $g_{2m} \sim g_m^2$ and $\mu_{2m} \sim g_{2m} \sim g_m^2$. Then we conclude that
\begin{align}
 & k\partial_k Y_0 \sim \Omega^{-1} g_1^3 \, , \nonumber \\
& k \partial_k X_0 = -\frac{\I S_d k^d \gamma}{\left|\text{Im}(M_0)\right|}\frac{Y_0}{2} + \mathcal{O}(\Omega^{-1} g_1^4) \, .
\end{align}
Neglecting the terms that are not $\mathcal{O}(\Omega^{-1})$ and $1$-loop ($\epsilon$ expansion) produces a closed set of equations
\begin{align}
 & k \partial_k \mu_0 = -\frac{S_d k^d \gamma}{\left|\text{Im}(M_0)\right|} \left(g_0 + \I X_0 \right) \, , \nonumber \\
& k \partial_k g_{0} = \frac{S_d k^d \gamma}{\left|\text{Im}(M_0)\right|} \nonumber \\
& \quad \times \left\{ \left[\frac{g_0}{2 M_0} +  \frac{g_0-(g_{0})^*}{\I  \text{Im}(M_0)} \right] \left(g_0 + \I X_0\right) + Y_{0} \right\} \, ,\nonumber \\
& k\partial_k Y_0 = 0 \, , \nonumber \\
& k \partial_k X_0 = -\frac{\I S_d k^d \gamma}{\left|\text{Im}(M_0)\right|}\frac{Y_0}{2}\, .
\label{eq_monochromatic_drive}
\end{align}

\subsection{Imaginary couplings}

\Eq{eq_monochromatic_drive} can be further simplified in the case of purely imaginary couplings. When the couplings can be written as $K = \I K'$, $\mu(t) = \I \mu'(t)$ and $g(t) = \I g'(t)$ with $K'$, $\mu'$ and $g'$ real, the Fourier modes satisfy
\begin{align}
  \mu_{-n} = -\mu_n^* \, , && g_{-n} = -g_n^* \, .
\end{align}
Inserting this in \Eq{eq_defXY} provides $Y_0 = 0$ and
\begin{align}
 X_0 = \frac{4}{\Omega} \sum_{m=1}^\infty \frac{\text{Im}(g_m^* \mu_m)}{m}\, ,
 \label{eq_defX_app}
\end{align}
which is a real number. The drive parameter $x$ is defined as the real part of $X_0$, and both are equal when the couplings are purely imaginary. The flow equations of the main text are written in terms of the imaginary couplings. Inserting ${\mu_0' = -\I \mu_0}$, ${g_0' = -\I g_0}$ and ${K' = -\I K}$ into \Eq{eq_dimensionful_flow_equations_0} with $Y_0 = 0$ and ${X_0 = \sum_m \tilde{g}_m}$ and rescaling $\mu_0'$ and $g_0'$ as $\mu_0$ and $g_0$ [see \Eq{eq_rescaling}] provides \Eqs{eq_dimensionless_rg_flow} and \eq{eq_x_flow} (with the primes dropped).

We have checked from \Eq{eq_dimensionful_flow_equations} that, in the case of purely imaginary couplings, no real parts are generated. We can safely assume that when $\mu$ and $g$ contain no real parts, they remain imaginary at all scales. This was already observed in \cite{Sieberer2013a} for time-independent couplings and is related to an additional symmetry that emerges for imaginary couplings.

\section{Equilibrium symmetry}
\label{app_equilibrium_symmetry}

The presence of thermal equilibrium can be framed in terms of a microscopic symmetry of the dynamic action \cite{Sieberer2013b,tauber2014critical}. In this section, we define the corresponding field transformation and (along the lines of \cite{Sieberer2013b}) discuss the conditions under which it is a symmetry of our system (\ie when the dynamical action describes a thermally equilibrated system). It is instructive to consider a more general action
\begin{align*}
 S = \int_{t,\boldsymbol{x}} \tilde{\phi}^* & \left[ Z^* \I \partial_t \phi - \mathcal{K}(\left|\phi\right|,t) \phi  \right]  + \text{c.c.} + \I \gamma |\tilde{\phi}|^2 \, .
\end{align*}
In principle, all the parameters in the above action are complex periodic functions of time. The interaction as well as the kinetic term are bundled in the operator ${\mathcal{K}(\left|\phi\right|,t) = K \nabla^2 - \mu - g\left|\phi\right|^2}$. Then we define the field transformation
\begin{align}
 \left(\begin{array}{c}
        \phi(t) \\ \tilde{\phi}(t) \\ \phi^*(t) \\ \tilde{\phi}^*(t)
       \end{array} \right) \rightarrow \left( \hspace{-3pt} \begin{array}{c} \phi^*(-t) \\
       \frac{r-\I}{r+\I} \tilde{\phi}^*(-t) + \frac{r-\I}{\text{Re}(Z)- r \text{Im}(Z)} \frac{\I}{2 T} \partial_t \phi^*(-t) \\
       \phi(-t) \\
       \frac{r+\I}{r-\I} \tilde{\phi}(-t) + \frac{r+\I}{\text{Re}(Z)- r \text{Im}(Z)} \frac{\I}{2 T} \partial_t \phi(-t)\end{array} \hspace{-3pt} \right) ,
       \label{eq_equilibrium_symmetry}
\end{align}
with two (yet unspecified) parameters, $r$ and $T$. This transformation is called 'equilibrium symmetry' because, if it is possible to find values of $r$ and $T$ such that the above transformation is a symmetry of the action, then it can be shown \cite{Sieberer2015,Aron2018} that the system obeys fluctuation-dissipation relations with a temperature given by $T$.

We find that our driven system is at equilibrium when:
\begin{enumerate}[(i)]
 \item All the time-dependent couplings are even in $t$ (up to a global time shift).

 \item There is a single (possibly time-dependent) real number $r(t)$ such that
  \begin{align}
    \text{Re}(\mathcal{K}) = - r \text{Im}(\mathcal{K}) \, .
  \end{align}
  This is a generalization of the requirement (found in \cite{Sieberer2013b}) that all the couplings lay on the same ray of the complex plane.

\item The time-dependence of $Z$, $\gamma$ and $r$ are such that the temperature
\begin{align}
 T = \frac{(1+r^2) \tilde{\gamma}}{4[\text{Re}(Z)-r \text{Im}(Z)]} \, ,
\end{align}
with ${\tilde{\gamma} = {\gamma}/[\text{Re}(Z)-r \text{Im}(Z)}]$, does not depend on time. This relation also defines $T$.

\item $\mathcal{K}$ satisfies
\begin{align}
 \frac{\partial}{\partial t} \left[ \frac{1}{\tilde{\gamma}} \text{Im}[\mathcal{K}(\left|\phi\right|,t)]\right] = 0 \, .
\end{align}
The time derivative does not hit the field in the above equation. This equation must be valid for all values of $\phi$. It implies that there exists a time-independent operator $\mathcal{K}(\left|\phi\right|)$ such that
\begin{align}
 \mathcal{K}(\left|\phi\right|,t) = \tilde{\gamma} \, \mathcal{K}(\left|\phi\right|) \, .
\end{align}
\Ie the time-dependent couplings oscillate in phase with each other.
\end{enumerate}

\section{Physical interpretation of \texorpdfstring{$x$}{x}}
\label{app_x}

In this section we show how the value of $x$ \Eqs{eq_defX_app} and \eq{eq:xxx} can be related to the phases of $\mu_1$ and $g_1$. Moreover, we argue that $x$ being identically zero can only happen at equilibrium. We focus on a monochromatic drive and purely imaginary $\mu(t)$ and $g(t)$,
\begin{align}
 & \mu_1 = |\mu_1|\text{e}^{-\I\theta_{\mu}} \, , && \mu_{-1} = -|\mu_1|\text{e}^{\I\theta_{\mu}} \, , \nonumber \\
 & g_1=|g_1|\text{e}^{-\I\theta_{g}} \, , && g_{-1}=-|g_1|\text{e}^{\I\theta_{g}} \, .
\end{align}

We start by giving a meaning to the phases of the couplings, by inserting them back into \Eq{eq_def_mn_gn},
\begin{align*}
& \mu - \mu_0 = \mu_1 \, \text{e}^{-\I \Omega t} - \mu_{1}^* \, \text{e}^{\I \Omega t} = 2 \I |\mu_1| \sin( \Omega t+\theta_{\mu}) \, , \nonumber \\
& g - g_0 = g_1 \, \text{e}^{-\I \Omega t} - g_1^* \, \text{e}^{\I \Omega t} = 2\I |g_1| \sin(\Omega t+\theta_g) \, .
\end{align*}
The phases of the drive couplings are phases in the time dependence of $\mu$ and $g$.

Inserting $\mu_1 = |\mu_1|\text{e}^{-\I\theta_{\mu}}$ and $g_1=|g_1|\text{e}^{-\I\theta_{g}}$ into \Eq{eq_defX_app} provides,
\begin{align}\label{eq:xxx}
 X_0 = 4\frac{|\mu_1| |g_1| \sin(\theta_g-\theta_\mu)}{\Omega} \, , \, X_0 = \text{Re}(X_0) \equiv x.
\end{align}
This shows that the relative phase of $\mu$ and $g$ plays an important role. Indeed, $x$ can be made to vanish when $\mu$ and $g$ are proportional to each other. On the other hand, $|x|$ is amplified when one of the couplings is lagging behind the other by a quarter of the drive period.

The present calculation may suggest that our effect is absent in a driven system when either $\mu_1$ or $g_1$ is zero, or when $\theta_g = \theta_{\mu}$. This is however only true at $\mathcal{O}(\Omega^{-1})$. If $x=0$ in a driven microscopic model, a non-vanishing value of $x$ will inevitably build up under renormalization, however, only at $\mathcal \mathcal{O}(\Omega^{-2})$. This is inferred from the analysis of \Sect{app_equilibrium_symmetry}, which shows that $x$ vanishes identically only when the microscopic couplings are just right for \Eq{eq_equilibrium_symmetry} to be a symmetry of the dynamical action. This corresponds to an unnatural fine tuning of the parameters of the model.

The case of couplings with a real time-dependent part, which arises naturally when the underlying system Hamiltonian is time-dependent but the couplings to the bath are not, falls in the above category as well. Although, at a first glance, it looks like $\hat{x}=0$, a non vanishing $\hat{x}$ is generated by the \RG.

More generally, $x$ is built up from the following picture: We distinguish $3$ levels for the problem based on the scale of observation:
\begin{itemize}
 \item The time-dependent Hamiltoinan is defined at the {\bf microscopic} level. It is coupled to a bath. We do not resolve this level.
 \item Perturbative corrections in powers of $\Omega^{-1}$ occur at the {\bf mesoscopic} level. They provide \Eq{eq_ddgpe}, where the periodic drive and the dissipation are encoded in the complex, time-dependent couplings. This is the level at which we start our calculation.
 \item Phase transitions are visible at the {\bf macroscopic} level. In particular, the transition is first order there if there are time-dependent couplings with imaginary parts at the mesoscopic level.
\end{itemize}
We conclude that the phase transition is unavoidably first order because it would require fine-tuning for the ingredients (time-dependent couplings with imaginary parts) to be missing at the mesoscopic level. This is an \RG argument: All couplings that can be generated (\ie are compatible with the system's symmetries) will be. If these are not present at the microscopic level however, they will appear at $\mathcal{O}(\Omega^{-1})$ at the mesoscopic level and therefore be $\mathcal{O}(\Omega^{-2})$ at the macroscopic level.
The transition will be very weakly first-order.

\section{Far-from-equilibrium fixed point}
\label{app_farfromequilibrium}

In the present work we find that the inclusion of a periodic drive produces an additional relevant coupling which ultimately leads to a runaway flow. This implies that criticality is only visible if the drive parameter is tuned to zero $\hat{x}=0$. It is however possible that another attractive fixed point lies beyond the reach of our approximations. Then the system would be critical for a finite $\hat{x}$.

If there was such a fixed point, then there would be a cross-over from the equilibrium criticality to this new Floquet criticality. The cross-over scale would be given by the cut-off scale at which the \RG flow can reach the vicinity of this far-from-equilibrium fixed point. This would be a highly non-universal scale. In particular it would depend on the drive amplitude since this controls the initial conditions of the \RG flow and the distance (in \RG steps) to this new fixed point.

Although we can not exclude this eventuality, we find that if such a fixed point exists, then its coordinates ($g^*$, $\mu^*$, $x^*$, etc.) must lie far away from the \WF fixed point. Indeed, it would be obtained by balancing the $\mathcal{O}(\Omega^{-1})$ terms with terms of higher order in $\Omega^{-1}$. Then even though the large-scale physics would indeed be controlled by this fixed point, the cross-over scale would be very large. Based on dimensional analysis, we can expect the cross-over scale to behave as $k^*\sim \Omega^{-1/2}$. In a system with a finite size, the phase transition would effectively remain first order for $\Omega$ large enough.

\section{Analysis of the \texorpdfstring{\RG}{RG} fixed point}
\label{app_fixed_point}

In this section we give explicit expressions for the stability matrix and its eigensystem at the \WF fixed point, which emerges when $\hat{x}=0$ in \Eqs{eq_dimensionless_rg_flow} and \eq{eq_x_flow} (the canonical rescaling is ${\hat x = k^{d-4} {\gamma x}/[{4 \text{Im}(K)^2}]}$).

To order ${\epsilon = 4-d}$, the fixed point coordinates are
\begin{align}
\vec{g}^* = \left(\begin{array}{c} \mu^* \\ g^* \\ x^* \end{array} \right) \cong \left( \begin{array}{c} -\epsilon/5 \\ 4 \pi^2 \epsilon/5 \\ 0 \end{array} \right) \, .
\end{align}
The stability matrix is then obtained by computing the Jacobian matrix of the right-hand-side of the \RG flow equations and evaluating it at the \WF fixed point. We find
\begin{align*}
 M = \left(\begin{array}{ccc} 
 -2 + \frac{2\epsilon}{5} & -\frac{4S_d}{5}(5+\epsilon) & -\frac{4S_d}{5}(5+\epsilon) \\
            -\frac{\epsilon^2}{S_d(5+\epsilon)} & \epsilon & \epsilon \\
            0 & 0 & - \epsilon
           \end{array}\right) \, .
\end{align*}

The critical exponents are the eigenvalues of $M$. They are defined by
\begin{align}
 M \vec{v}_1 = -\frac{1}{\nu} \, \vec{v}_1 \, , && M \vec{v}_2 = \epsilon \, \vec{v}_2 \, , &&  M \vec{v}_3 = -\frac{1}{\nu_d} \, \vec{v}_3 \, ,
\end{align}
and are given by $1/\nu \cong 2 -2 \epsilon/5$ and $1/\nu_d = \epsilon$ [to $\mathcal{O}(\epsilon)$]. The corresponding eigenvectors are
\begin{align*}
 & \vec{v}_1 = \left( \begin{array}{c} \frac{5}{4\pi^2 \epsilon} + \frac{3+ 2P}{4\pi^2} + \frac{\epsilon}{4} \left(\frac{11+2P(P+3)}{5\pi^2}-\frac{5}{48}\right) \\ \epsilon \\ 0 \end{array}\right) \, , \nonumber \\
 & \vec{v}_2 = \left( \begin{array}{c} -\frac{1}{4\pi^2}-\frac{P \epsilon}{10 \pi^2}\\ 1 \\ 0 \end{array}\right) \, , \quad \vec{v}_3 = \left( \begin{array}{c} -\frac{1}{8\pi^2}-\frac{(P+2)\epsilon}{20\pi^2} \\ -\frac{1}{2}- \frac{\epsilon}{10} \\ 1 \end{array}\right) \, ,
\end{align*}
with ${P = [4-5(\gamma-\log(4\pi))]/4\cong 3.442}$, and $\gamma \cong 0.577$ is Euler's constant. The two negative critical exponents correspond to the two relevant directions, $\vec{v}_1$ and $\vec{v}_3$.

Writing ${(\delta \mu,\delta g,\hat{x})}$ as a linear combination of $\vec{v}_i$ and expanding the result to linear order in $\epsilon$ provides generic initial conditions for the \RG flow as
\begin{align*}
 (\delta\mu,\delta g,\hat{x}) = & \frac{\epsilon\left(\delta t + A \hat{x}\right)}{5A} \, \vec{v}_1 + \left[\delta g+\frac{\hat{x}}{2}(1+\frac{\epsilon}{5})\right] \vec{v}_2  + \hat{x} \, \vec{v}_3 \, ,
\end{align*}
with $\delta t = A (\delta g + 4 \pi^2 \delta \mu)$, $\delta \mu = \hat{\mu}_0-\mu_0^*$, $\delta g = \hat{g}_0-g_0^*$ and $A>0$ a non-universal constant.

The fixed point is attainable only when the projection of $(\delta \mu, \delta g,\hat{x})$ along $\vec{v}_1$ and $\vec{v}_3$ vanishes. If this is not the case, the sign of $\Delta t = \delta t + A \hat{x}$ determines the macroscopic phase. For $\Delta t>0$, the system is in the symmetric phase, and the $O(2)$ symmetry is broken for $\Delta t<0$. We see that a finite value of $\hat{x}$ shifts the location of the phase transition.

%


\begin{thebibliography}{88}%
\makeatletter
\providecommand \@ifxundefined [1]{%
 \@ifx{#1\undefined}
}%
\providecommand \@ifnum [1]{%
 \ifnum #1\expandafter \@firstoftwo
 \else \expandafter \@secondoftwo
 \fi
}%
\providecommand \@ifx [1]{%
 \ifx #1\expandafter \@firstoftwo
 \else \expandafter \@secondoftwo
 \fi
}%
\providecommand \natexlab [1]{#1}%
\providecommand \enquote  [1]{``#1''}%
\providecommand \bibnamefont  [1]{#1}%
\providecommand \bibfnamefont [1]{#1}%
\providecommand \citenamefont [1]{#1}%
\providecommand \href@noop [0]{\@secondoftwo}%
\providecommand \href [0]{\begingroup \@sanitize@url \@href}%
\providecommand \@href[1]{\@@startlink{#1}\@@href}%
\providecommand \@@href[1]{\endgroup#1\@@endlink}%
\providecommand \@sanitize@url [0]{\catcode `\\12\catcode `\$12\catcode
  `\&12\catcode `\#12\catcode `\^12\catcode `\_12\catcode `\%12\relax}%
\providecommand \@@startlink[1]{}%
\providecommand \@@endlink[0]{}%
\providecommand \url  [0]{\begingroup\@sanitize@url \@url }%
\providecommand \@url [1]{\endgroup\@href {#1}{\urlprefix }}%
\providecommand \urlprefix  [0]{URL }%
\providecommand \Eprint [0]{\href }%
\providecommand \doibase [0]{http://dx.doi.org/}%
\providecommand \selectlanguage [0]{\@gobble}%
\providecommand \bibinfo  [0]{\@secondoftwo}%
\providecommand \bibfield  [0]{\@secondoftwo}%
\providecommand \translation [1]{[#1]}%
\providecommand \BibitemOpen [0]{}%
\providecommand \bibitemStop [0]{}%
\providecommand \bibitemNoStop [0]{.\EOS\space}%
\providecommand \EOS [0]{\spacefactor3000\relax}%
\providecommand \BibitemShut  [1]{\csname bibitem#1\endcsname}%
\let\auto@bib@innerbib\@empty
\bibitem [{\citenamefont {Eckardt}(2017)}]{Eckardt2017}%
  \BibitemOpen
  \bibfield  {author} {\bibinfo {author} {\bibfnamefont {A.}~\bibnamefont
  {Eckardt}},\ }\href {\doibase 10.1103/RevModPhys.89.011004} {\bibfield
  {journal} {\bibinfo  {journal} {Rev. Mod. Phys.}\ }\textbf {\bibinfo {volume}
  {89}},\ \bibinfo {pages} {011004} (\bibinfo {year} {2017})},\ \Eprint
  {http://arxiv.org/abs/1606.08041} {arXiv:1606.08041 [cond-mat.quant-gas]}
  \BibitemShut {NoStop}%
\bibitem [{\citenamefont {{Moessner}}\ and\ \citenamefont
  {{Sondhi}}(2017)}]{Moessner2017}%
  \BibitemOpen
  \bibfield  {author} {\bibinfo {author} {\bibfnamefont {R.}~\bibnamefont
  {{Moessner}}}\ and\ \bibinfo {author} {\bibfnamefont {S.~L.}\ \bibnamefont
  {{Sondhi}}},\ }\href {\doibase 10.1038/nphys4106} {\bibfield  {journal}
  {\bibinfo  {journal} {Nat. Phys.}\ }\textbf {\bibinfo {volume} {13}},\
  \bibinfo {pages} {424} (\bibinfo {year} {2017})},\ \Eprint
  {http://arxiv.org/abs/1701.08056} {arXiv:1701.08056 [cond-mat.dis-nn]}
  \BibitemShut {NoStop}%
\bibitem [{\citenamefont {Struck}\ \emph {et~al.}(2013)\citenamefont {Struck},
  \citenamefont {Weinberg}, \citenamefont {\"{O}lschl\"{a}ger}, \citenamefont
  {Windpassinger}, \citenamefont {Simonet}, \citenamefont {Sengstock},
  \citenamefont {Hoppner}, \citenamefont {Hauke}, \citenamefont {Eckardt},
  \citenamefont {Lewenstein},\ and\ \citenamefont {Mathey}}]{Struck2013}%
  \BibitemOpen
  \bibfield  {author} {\bibinfo {author} {\bibfnamefont {J.}~\bibnamefont
  {Struck}}, \bibinfo {author} {\bibfnamefont {M.}~\bibnamefont {Weinberg}},
  \bibinfo {author} {\bibfnamefont {C.}~\bibnamefont {\"{O}lschl\"{a}ger}},
  \bibinfo {author} {\bibfnamefont {P.}~\bibnamefont {Windpassinger}}, \bibinfo
  {author} {\bibfnamefont {J.}~\bibnamefont {Simonet}}, \bibinfo {author}
  {\bibfnamefont {K.}~\bibnamefont {Sengstock}}, \bibinfo {author}
  {\bibfnamefont {R.}~\bibnamefont {Hoppner}}, \bibinfo {author} {\bibfnamefont
  {P.}~\bibnamefont {Hauke}}, \bibinfo {author} {\bibfnamefont
  {A.}~\bibnamefont {Eckardt}}, \bibinfo {author} {\bibfnamefont
  {M.}~\bibnamefont {Lewenstein}}, \ and\ \bibinfo {author} {\bibfnamefont
  {L.}~\bibnamefont {Mathey}},\ }\href {\doibase 10.1038/nphys2750} {\bibfield
  {journal} {\bibinfo  {journal} {Nat. Phys.}\ }\textbf {\bibinfo {volume}
  {9}},\ \bibinfo {pages} {738} (\bibinfo {year} {2013})},\ \Eprint
  {http://arxiv.org/abs/1304.5520} {arXiv:1304.5520 [cond-mat.quant-gas]}
  \BibitemShut {NoStop}%
\bibitem [{\citenamefont {Goldman}\ and\ \citenamefont
  {Dalibard}(2014)}]{Goldman2014}%
  \BibitemOpen
  \bibfield  {author} {\bibinfo {author} {\bibfnamefont {N.}~\bibnamefont
  {Goldman}}\ and\ \bibinfo {author} {\bibfnamefont {J.}~\bibnamefont
  {Dalibard}},\ }\href {\doibase 10.1103/PhysRevX.4.031027} {\bibfield
  {journal} {\bibinfo  {journal} {Phys. Rev. X}\ }\textbf {\bibinfo {volume}
  {4}},\ \bibinfo {pages} {031027} (\bibinfo {year} {2014})},\ \Eprint
  {http://arxiv.org/abs/1404.4373} {arXiv:1404.4373 [cond-mat.quant-gas]}
  \BibitemShut {NoStop}%
\bibitem [{\citenamefont {{Choi}}\ \emph {et~al.}(2017)\citenamefont {{Choi}},
  \citenamefont {{Choi}}, \citenamefont {{Landig}}, \citenamefont {{Kucsko}},
  \citenamefont {{Zhou}}, \citenamefont {{Isoya}}, \citenamefont {{Jelezko}},
  \citenamefont {{Onoda}}, \citenamefont {{Sumiya}}, \citenamefont {{Khemani}},
  \citenamefont {{von Keyserlingk}}, \citenamefont {{Yao}}, \citenamefont
  {{Demler}},\ and\ \citenamefont {{Lukin}}}]{Choi2017}%
  \BibitemOpen
  \bibfield  {author} {\bibinfo {author} {\bibfnamefont {S.}~\bibnamefont
  {{Choi}}}, \bibinfo {author} {\bibfnamefont {J.}~\bibnamefont {{Choi}}},
  \bibinfo {author} {\bibfnamefont {R.}~\bibnamefont {{Landig}}}, \bibinfo
  {author} {\bibfnamefont {G.}~\bibnamefont {{Kucsko}}}, \bibinfo {author}
  {\bibfnamefont {H.}~\bibnamefont {{Zhou}}}, \bibinfo {author} {\bibfnamefont
  {J.}~\bibnamefont {{Isoya}}}, \bibinfo {author} {\bibfnamefont
  {F.}~\bibnamefont {{Jelezko}}}, \bibinfo {author} {\bibfnamefont
  {S.}~\bibnamefont {{Onoda}}}, \bibinfo {author} {\bibfnamefont
  {H.}~\bibnamefont {{Sumiya}}}, \bibinfo {author} {\bibfnamefont
  {V.}~\bibnamefont {{Khemani}}}, \bibinfo {author} {\bibfnamefont
  {C.}~\bibnamefont {{von Keyserlingk}}}, \bibinfo {author} {\bibfnamefont
  {N.~Y.}\ \bibnamefont {{Yao}}}, \bibinfo {author} {\bibfnamefont
  {E.}~\bibnamefont {{Demler}}}, \ and\ \bibinfo {author} {\bibfnamefont
  {M.~D.}\ \bibnamefont {{Lukin}}},\ }\href {\doibase 10.1038/nature21426}
  {\bibfield  {journal} {\bibinfo  {journal} {Nature}\ }\textbf {\bibinfo
  {volume} {543}},\ \bibinfo {pages} {221} (\bibinfo {year} {2017})},\ \Eprint
  {http://arxiv.org/abs/1610.08057} {arXiv:1610.08057 [quant-ph]} \BibitemShut
  {NoStop}%
\bibitem [{\citenamefont {{Zhang}}\ \emph {et~al.}(2017)\citenamefont
  {{Zhang}}, \citenamefont {{Hess}}, \citenamefont {{Kyprianidis}},
  \citenamefont {{Becker}}, \citenamefont {{Lee}}, \citenamefont {{Smith}},
  \citenamefont {{Pagano}}, \citenamefont {{Potirniche}}, \citenamefont
  {{Potter}}, \citenamefont {{Vishwanath}}, \citenamefont {{Yao}},\ and\
  \citenamefont {{Monroe}}}]{Zhang2017}%
  \BibitemOpen
  \bibfield  {author} {\bibinfo {author} {\bibfnamefont {J.}~\bibnamefont
  {{Zhang}}}, \bibinfo {author} {\bibfnamefont {P.~W.}\ \bibnamefont {{Hess}}},
  \bibinfo {author} {\bibfnamefont {A.}~\bibnamefont {{Kyprianidis}}}, \bibinfo
  {author} {\bibfnamefont {P.}~\bibnamefont {{Becker}}}, \bibinfo {author}
  {\bibfnamefont {A.}~\bibnamefont {{Lee}}}, \bibinfo {author} {\bibfnamefont
  {J.}~\bibnamefont {{Smith}}}, \bibinfo {author} {\bibfnamefont
  {G.}~\bibnamefont {{Pagano}}}, \bibinfo {author} {\bibfnamefont {I.-D.}\
  \bibnamefont {{Potirniche}}}, \bibinfo {author} {\bibfnamefont {A.~C.}\
  \bibnamefont {{Potter}}}, \bibinfo {author} {\bibfnamefont {A.}~\bibnamefont
  {{Vishwanath}}}, \bibinfo {author} {\bibfnamefont {N.~Y.}\ \bibnamefont
  {{Yao}}}, \ and\ \bibinfo {author} {\bibfnamefont {C.}~\bibnamefont
  {{Monroe}}},\ }\href {\doibase 10.1038/nature21413} {\bibfield  {journal}
  {\bibinfo  {journal} {Nature}\ }\textbf {\bibinfo {volume} {543}},\ \bibinfo
  {pages} {217} (\bibinfo {year} {2017})},\ \Eprint
  {http://arxiv.org/abs/1609.08684} {arXiv:1609.08684 [quant-ph]} \BibitemShut
  {NoStop}%
\bibitem [{\citenamefont {D'Alessio}\ and\ \citenamefont
  {Rigol}(2014)}]{DAlessio2014}%
  \BibitemOpen
  \bibfield  {author} {\bibinfo {author} {\bibfnamefont {L.}~\bibnamefont
  {D'Alessio}}\ and\ \bibinfo {author} {\bibfnamefont {M.}~\bibnamefont
  {Rigol}},\ }\href {\doibase 10.1103/PhysRevX.4.041048} {\bibfield  {journal}
  {\bibinfo  {journal} {Phys. Rev. X}\ }\textbf {\bibinfo {volume} {4}},\
  \bibinfo {pages} {041048} (\bibinfo {year} {2014})},\ \Eprint
  {http://arxiv.org/abs/1402.5141} {arXiv:1402.5141 [cond-mat.stat-mech]}
  \BibitemShut {NoStop}%
\bibitem [{\citenamefont {Lazarides}\ \emph {et~al.}(2014)\citenamefont
  {Lazarides}, \citenamefont {Das},\ and\ \citenamefont
  {Moessner}}]{Lazarides2014}%
  \BibitemOpen
  \bibfield  {author} {\bibinfo {author} {\bibfnamefont {A.}~\bibnamefont
  {Lazarides}}, \bibinfo {author} {\bibfnamefont {A.}~\bibnamefont {Das}}, \
  and\ \bibinfo {author} {\bibfnamefont {R.}~\bibnamefont {Moessner}},\ }\href
  {\doibase 10.1103/PhysRevE.90.012110} {\bibfield  {journal} {\bibinfo
  {journal} {Phys. Rev. E}\ }\textbf {\bibinfo {volume} {90}},\ \bibinfo
  {pages} {012110} (\bibinfo {year} {2014})},\ \Eprint
  {http://arxiv.org/abs/1403.2946} {arXiv:1403.2946 [cond-mat.stat-mech]}
  \BibitemShut {NoStop}%
\bibitem [{\citenamefont {Shirai}\ \emph {et~al.}(2015)\citenamefont {Shirai},
  \citenamefont {Mori},\ and\ \citenamefont {Miyashita}}]{Shirai2015}%
  \BibitemOpen
  \bibfield  {author} {\bibinfo {author} {\bibfnamefont {T.}~\bibnamefont
  {Shirai}}, \bibinfo {author} {\bibfnamefont {T.}~\bibnamefont {Mori}}, \ and\
  \bibinfo {author} {\bibfnamefont {S.}~\bibnamefont {Miyashita}},\ }\href
  {\doibase 10.1103/PhysRevE.91.030101} {\bibfield  {journal} {\bibinfo
  {journal} {Phys. Rev. E}\ }\textbf {\bibinfo {volume} {91}},\ \bibinfo
  {pages} {030101} (\bibinfo {year} {2015})},\ \Eprint
  {http://arxiv.org/abs/1410.0464} {arXiv:1410.0464 [cond-mat.stat-mech]}
  \BibitemShut {NoStop}%
\bibitem [{\citenamefont {Chandran}\ and\ \citenamefont
  {Sondhi}(2016)}]{Chandran2016}%
  \BibitemOpen
  \bibfield  {author} {\bibinfo {author} {\bibfnamefont {A.}~\bibnamefont
  {Chandran}}\ and\ \bibinfo {author} {\bibfnamefont {S.~L.}\ \bibnamefont
  {Sondhi}},\ }\href {\doibase 10.1103/PhysRevB.93.174305} {\bibfield
  {journal} {\bibinfo  {journal} {Phys. Rev. B}\ }\textbf {\bibinfo {volume}
  {93}},\ \bibinfo {pages} {174305} (\bibinfo {year} {2016})},\ \Eprint
  {http://arxiv.org/abs/1506.08836} {arXiv:1506.08836 [cond-mat.stat-mech]}
  \BibitemShut {NoStop}%
\bibitem [{\citenamefont {Canovi}\ \emph {et~al.}(2016)\citenamefont {Canovi},
  \citenamefont {Kollar},\ and\ \citenamefont {Eckstein}}]{Canovi2016}%
  \BibitemOpen
  \bibfield  {author} {\bibinfo {author} {\bibfnamefont {E.}~\bibnamefont
  {Canovi}}, \bibinfo {author} {\bibfnamefont {M.}~\bibnamefont {Kollar}}, \
  and\ \bibinfo {author} {\bibfnamefont {M.}~\bibnamefont {Eckstein}},\ }\href
  {\doibase 10.1103/PhysRevE.93.012130} {\bibfield  {journal} {\bibinfo
  {journal} {Phys. Rev. E}\ }\textbf {\bibinfo {volume} {93}},\ \bibinfo
  {pages} {012130} (\bibinfo {year} {2016})},\ \Eprint
  {http://arxiv.org/abs/1507.00991} {arXiv:1507.00991 [cond-mat.str-el]}
  \BibitemShut {NoStop}%
\bibitem [{\citenamefont {Bukov}\ \emph {et~al.}(2015)\citenamefont {Bukov},
  \citenamefont {Gopalakrishnan}, \citenamefont {Knap},\ and\ \citenamefont
  {Demler}}]{Bukov2015b}%
  \BibitemOpen
  \bibfield  {author} {\bibinfo {author} {\bibfnamefont {M.}~\bibnamefont
  {Bukov}}, \bibinfo {author} {\bibfnamefont {S.}~\bibnamefont
  {Gopalakrishnan}}, \bibinfo {author} {\bibfnamefont {M.}~\bibnamefont
  {Knap}}, \ and\ \bibinfo {author} {\bibfnamefont {E.}~\bibnamefont
  {Demler}},\ }\href {\doibase 10.1103/PhysRevLett.115.205301} {\bibfield
  {journal} {\bibinfo  {journal} {Phys. Rev. Lett.}\ }\textbf {\bibinfo
  {volume} {115}},\ \bibinfo {pages} {205301} (\bibinfo {year} {2015})},\
  \Eprint {http://arxiv.org/abs/1507.01946} {arXiv:1507.01946
  [cond-mat.quant-gas]} \BibitemShut {NoStop}%
\bibitem [{\citenamefont {Khemani}\ \emph {et~al.}(2016)\citenamefont
  {Khemani}, \citenamefont {Lazarides}, \citenamefont {Moessner},\ and\
  \citenamefont {Sondhi}}]{Khemani2016}%
  \BibitemOpen
  \bibfield  {author} {\bibinfo {author} {\bibfnamefont {V.}~\bibnamefont
  {Khemani}}, \bibinfo {author} {\bibfnamefont {A.}~\bibnamefont {Lazarides}},
  \bibinfo {author} {\bibfnamefont {R.}~\bibnamefont {Moessner}}, \ and\
  \bibinfo {author} {\bibfnamefont {S.~L.}\ \bibnamefont {Sondhi}},\ }\href
  {\doibase 10.1103/PhysRevLett.116.250401} {\bibfield  {journal} {\bibinfo
  {journal} {Phys. Rev. Lett.}\ }\textbf {\bibinfo {volume} {116}},\ \bibinfo
  {pages} {250401} (\bibinfo {year} {2016})},\ \Eprint
  {http://arxiv.org/abs/1508.03344} {arXiv:1508.03344 [cond-mat.dis-nn]}
  \BibitemShut {NoStop}%
\bibitem [{\citenamefont {Genske}\ and\ \citenamefont
  {Rosch}(2015)}]{Genske2015}%
  \BibitemOpen
  \bibfield  {author} {\bibinfo {author} {\bibfnamefont {M.}~\bibnamefont
  {Genske}}\ and\ \bibinfo {author} {\bibfnamefont {A.}~\bibnamefont {Rosch}},\
  }\href {\doibase 10.1103/PhysRevA.92.062108} {\bibfield  {journal} {\bibinfo
  {journal} {Phys. Rev. A}\ }\textbf {\bibinfo {volume} {92}},\ \bibinfo
  {pages} {062108} (\bibinfo {year} {2015})},\ \Eprint
  {http://arxiv.org/abs/1508.04551} {arXiv:1508.04551 [cond-mat.quant-gas]}
  \BibitemShut {NoStop}%
\bibitem [{\citenamefont {Kuwahara}\ \emph {et~al.}(2016)\citenamefont
  {Kuwahara}, \citenamefont {Mori},\ and\ \citenamefont
  {Saito}}]{Kuwahara2016}%
  \BibitemOpen
  \bibfield  {author} {\bibinfo {author} {\bibfnamefont {T.}~\bibnamefont
  {Kuwahara}}, \bibinfo {author} {\bibfnamefont {T.}~\bibnamefont {Mori}}, \
  and\ \bibinfo {author} {\bibfnamefont {K.}~\bibnamefont {Saito}},\ }\href
  {\doibase 10.1016/j.aop.2016.01.012} {\bibfield  {journal} {\bibinfo
  {journal} {Ann. Phys.}\ }\textbf {\bibinfo {volume} {367}},\ \bibinfo {pages}
  {96 } (\bibinfo {year} {2016})},\ \Eprint {http://arxiv.org/abs/1508.05797}
  {arXiv:1508.05797 [quant-ph]} \BibitemShut {NoStop}%
\bibitem [{\citenamefont {Mori}\ \emph {et~al.}(2016)\citenamefont {Mori},
  \citenamefont {Kuwahara},\ and\ \citenamefont {Saito}}]{Mori2016}%
  \BibitemOpen
  \bibfield  {author} {\bibinfo {author} {\bibfnamefont {T.}~\bibnamefont
  {Mori}}, \bibinfo {author} {\bibfnamefont {T.}~\bibnamefont {Kuwahara}}, \
  and\ \bibinfo {author} {\bibfnamefont {K.}~\bibnamefont {Saito}},\ }\href
  {\doibase 10.1103/PhysRevLett.116.120401} {\bibfield  {journal} {\bibinfo
  {journal} {Phys. Rev. Lett.}\ }\textbf {\bibinfo {volume} {116}},\ \bibinfo
  {pages} {120401} (\bibinfo {year} {2016})},\ \Eprint
  {http://arxiv.org/abs/1509.03968} {arXiv:1509.03968 [cond-mat.stat-mech]}
  \BibitemShut {NoStop}%
\bibitem [{\citenamefont {Shirai}\ \emph {et~al.}(2016)\citenamefont {Shirai},
  \citenamefont {Thingna}, \citenamefont {Mori}, \citenamefont {Denisov},
  \citenamefont {H\"anggi},\ and\ \citenamefont {Miyashita}}]{Shirai2016}%
  \BibitemOpen
  \bibfield  {author} {\bibinfo {author} {\bibfnamefont {T.}~\bibnamefont
  {Shirai}}, \bibinfo {author} {\bibfnamefont {J.}~\bibnamefont {Thingna}},
  \bibinfo {author} {\bibfnamefont {T.}~\bibnamefont {Mori}}, \bibinfo {author}
  {\bibfnamefont {S.}~\bibnamefont {Denisov}}, \bibinfo {author} {\bibfnamefont
  {P.}~\bibnamefont {H\"anggi}}, \ and\ \bibinfo {author} {\bibfnamefont
  {S.}~\bibnamefont {Miyashita}},\ }\href {\doibase
  10.1088/1367-2630/18/5/053008} {\bibfield  {journal} {\bibinfo  {journal}
  {New J. Phys.}\ }\textbf {\bibinfo {volume} {18}},\ \bibinfo {pages} {053008}
  (\bibinfo {year} {2016})},\ \Eprint {http://arxiv.org/abs/1511.06864}
  {arXiv:1511.06864 [cond-mat.stat-mech]} \BibitemShut {NoStop}%
\bibitem [{\citenamefont {{Weidinger}}\ and\ \citenamefont
  {{Knap}}(2017)}]{Weidinger2017}%
  \BibitemOpen
  \bibfield  {author} {\bibinfo {author} {\bibfnamefont {S.~A.}\ \bibnamefont
  {{Weidinger}}}\ and\ \bibinfo {author} {\bibfnamefont {M.}~\bibnamefont
  {{Knap}}},\ }\href {\doibase 10.1038/srep45382} {\bibfield  {journal}
  {\bibinfo  {journal} {Sci. Rep.}\ }\textbf {\bibinfo {volume} {7}},\ \bibinfo
  {eid} {45382} (\bibinfo {year} {2017})},\ \Eprint
  {http://arxiv.org/abs/1609.09089} {arXiv:1609.09089 [cond-mat.quant-gas]}
  \BibitemShut {NoStop}%
\bibitem [{\citenamefont {Shirai}\ \emph {et~al.}(2018)\citenamefont {Shirai},
  \citenamefont {Mori},\ and\ \citenamefont {Miyashita}}]{Shirai2018}%
  \BibitemOpen
  \bibfield  {author} {\bibinfo {author} {\bibfnamefont {T.}~\bibnamefont
  {Shirai}}, \bibinfo {author} {\bibfnamefont {T.}~\bibnamefont {Mori}}, \ and\
  \bibinfo {author} {\bibfnamefont {S.}~\bibnamefont {Miyashita}},\ }\href
  {\doibase 10.1140/epjst/e2018-00087-1} {\bibfield  {journal} {\bibinfo
  {journal} {E. Phys. J. ST}\ }\textbf {\bibinfo {volume} {227}},\ \bibinfo
  {pages} {323} (\bibinfo {year} {2018})},\ \Eprint
  {http://arxiv.org/abs/1801.02838} {arXiv:1801.02838 [cond-mat.stat-mech]}
  \BibitemShut {NoStop}%
\bibitem [{\citenamefont {Howell}\ \emph {et~al.}(2019)\citenamefont {Howell},
  \citenamefont {Weinberg}, \citenamefont {Sels}, \citenamefont {Polkovnikov},\
  and\ \citenamefont {Bukov}}]{Howell2018}%
  \BibitemOpen
  \bibfield  {author} {\bibinfo {author} {\bibfnamefont {O.}~\bibnamefont
  {Howell}}, \bibinfo {author} {\bibfnamefont {P.}~\bibnamefont {Weinberg}},
  \bibinfo {author} {\bibfnamefont {D.}~\bibnamefont {Sels}}, \bibinfo {author}
  {\bibfnamefont {A.}~\bibnamefont {Polkovnikov}}, \ and\ \bibinfo {author}
  {\bibfnamefont {M.}~\bibnamefont {Bukov}},\ }\href {\doibase
  10.1103/PhysRevLett.122.010602} {\bibfield  {journal} {\bibinfo  {journal}
  {Phys. Rev. Lett.}\ }\textbf {\bibinfo {volume} {122}},\ \bibinfo {pages}
  {010602} (\bibinfo {year} {2019})},\ \Eprint
  {http://arxiv.org/abs/1802.04910} {arXiv:1802.04910 [cond-mat.stat-mech]}
  \BibitemShut {NoStop}%
\bibitem [{\citenamefont {Kitagawa}\ \emph {et~al.}(2010)\citenamefont
  {Kitagawa}, \citenamefont {Berg}, \citenamefont {Rudner},\ and\ \citenamefont
  {Demler}}]{Kitagawa10}%
  \BibitemOpen
  \bibfield  {author} {\bibinfo {author} {\bibfnamefont {T.}~\bibnamefont
  {Kitagawa}}, \bibinfo {author} {\bibfnamefont {E.}~\bibnamefont {Berg}},
  \bibinfo {author} {\bibfnamefont {M.}~\bibnamefont {Rudner}}, \ and\ \bibinfo
  {author} {\bibfnamefont {E.}~\bibnamefont {Demler}},\ }\href {\doibase
  10.1103/PhysRevB.82.235114} {\bibfield  {journal} {\bibinfo  {journal} {Phys.
  Rev. B}\ }\textbf {\bibinfo {volume} {82}},\ \bibinfo {pages} {235114}
  (\bibinfo {year} {2010})},\ \Eprint {http://arxiv.org/abs/1010.6126}
  {arXiv:1010.6126 [cond-mat.mes-hall]} \BibitemShut {NoStop}%
\bibitem [{\citenamefont {{Lindner}}\ \emph {et~al.}(2011)\citenamefont
  {{Lindner}}, \citenamefont {{Refael}},\ and\ \citenamefont
  {{Galitski}}}]{Lindner2011}%
  \BibitemOpen
  \bibfield  {author} {\bibinfo {author} {\bibfnamefont {N.~H.}\ \bibnamefont
  {{Lindner}}}, \bibinfo {author} {\bibfnamefont {G.}~\bibnamefont {{Refael}}},
  \ and\ \bibinfo {author} {\bibfnamefont {V.}~\bibnamefont {{Galitski}}},\
  }\href {\doibase 10.1038/nphys1926} {\bibfield  {journal} {\bibinfo
  {journal} {Nat. Phys.}\ }\textbf {\bibinfo {volume} {7}},\ \bibinfo {pages}
  {490} (\bibinfo {year} {2011})},\ \Eprint {http://arxiv.org/abs/1008.1792}
  {arXiv:1008.1792 [cond-mat.mtrl-sci]} \BibitemShut {NoStop}%
\bibitem [{\citenamefont {{Cayssol}}\ \emph {et~al.}(2013)\citenamefont
  {{Cayssol}}, \citenamefont {{D{\'o}ra}}, \citenamefont {{Simon}},\ and\
  \citenamefont {{Moessner}}}]{Cayssol2012}%
  \BibitemOpen
  \bibfield  {author} {\bibinfo {author} {\bibfnamefont {J.}~\bibnamefont
  {{Cayssol}}}, \bibinfo {author} {\bibfnamefont {B.}~\bibnamefont
  {{D{\'o}ra}}}, \bibinfo {author} {\bibfnamefont {F.}~\bibnamefont {{Simon}}},
  \ and\ \bibinfo {author} {\bibfnamefont {R.}~\bibnamefont {{Moessner}}},\
  }\href {\doibase 10.1002/pssr.201206451} {\bibfield  {journal} {\bibinfo
  {journal} {Phys. Status Solidi RRL}\ }\textbf {\bibinfo {volume} {7}},\
  \bibinfo {pages} {101} (\bibinfo {year} {2013})},\ \Eprint
  {http://arxiv.org/abs/1211.5623} {arXiv:1211.5623 [cond-mat.mes-hall]}
  \BibitemShut {NoStop}%
\bibitem [{\citenamefont {Rudner}\ \emph {et~al.}(2013)\citenamefont {Rudner},
  \citenamefont {Lindner}, \citenamefont {Berg},\ and\ \citenamefont
  {Levin}}]{Rudner2013}%
  \BibitemOpen
  \bibfield  {author} {\bibinfo {author} {\bibfnamefont {M.~S.}\ \bibnamefont
  {Rudner}}, \bibinfo {author} {\bibfnamefont {N.~H.}\ \bibnamefont {Lindner}},
  \bibinfo {author} {\bibfnamefont {E.}~\bibnamefont {Berg}}, \ and\ \bibinfo
  {author} {\bibfnamefont {M.}~\bibnamefont {Levin}},\ }\href {\doibase
  10.1103/PhysRevX.3.031005} {\bibfield  {journal} {\bibinfo  {journal} {Phys.
  Rev. X}\ }\textbf {\bibinfo {volume} {3}},\ \bibinfo {pages} {031005}
  (\bibinfo {year} {2013})},\ \Eprint {http://arxiv.org/abs/1212.3324}
  {arXiv:1212.3324 [cond-mat.mes-hall]} \BibitemShut {NoStop}%
\bibitem [{\citenamefont {Karzig}\ \emph {et~al.}(2015)\citenamefont {Karzig},
  \citenamefont {Bardyn}, \citenamefont {Lindner},\ and\ \citenamefont
  {Refael}}]{Karzig2015}%
  \BibitemOpen
  \bibfield  {author} {\bibinfo {author} {\bibfnamefont {T.}~\bibnamefont
  {Karzig}}, \bibinfo {author} {\bibfnamefont {C.-E.}\ \bibnamefont {Bardyn}},
  \bibinfo {author} {\bibfnamefont {N.~H.}\ \bibnamefont {Lindner}}, \ and\
  \bibinfo {author} {\bibfnamefont {G.}~\bibnamefont {Refael}},\ }\href
  {\doibase 10.1103/PhysRevX.5.031001} {\bibfield  {journal} {\bibinfo
  {journal} {Phys. Rev. X}\ }\textbf {\bibinfo {volume} {5}},\ \bibinfo {pages}
  {031001} (\bibinfo {year} {2015})},\ \Eprint {http://arxiv.org/abs/1406.4156}
  {arXiv:1406.4156 [cond-mat.quant-gas]} \BibitemShut {NoStop}%
\bibitem [{\citenamefont {D'Alessio}\ and\ \citenamefont
  {Polkovnikov}(2013)}]{DAlessio2013}%
  \BibitemOpen
  \bibfield  {author} {\bibinfo {author} {\bibfnamefont {L.}~\bibnamefont
  {D'Alessio}}\ and\ \bibinfo {author} {\bibfnamefont {A.}~\bibnamefont
  {Polkovnikov}},\ }\href {\doibase 10.1016/j.aop.2013.02.011} {\bibfield
  {journal} {\bibinfo  {journal} {Ann. Phys.}\ }\textbf {\bibinfo {volume}
  {333}},\ \bibinfo {pages} {19 } (\bibinfo {year} {2013})},\ \Eprint
  {http://arxiv.org/abs/1210.2791} {arXiv:1210.2791 [cond-mat.stat-mech]}
  \BibitemShut {NoStop}%
\bibitem [{\citenamefont {Lazarides}\ \emph {et~al.}(2015)\citenamefont
  {Lazarides}, \citenamefont {Das},\ and\ \citenamefont
  {Moessner}}]{Lazarides2015}%
  \BibitemOpen
  \bibfield  {author} {\bibinfo {author} {\bibfnamefont {A.}~\bibnamefont
  {Lazarides}}, \bibinfo {author} {\bibfnamefont {A.}~\bibnamefont {Das}}, \
  and\ \bibinfo {author} {\bibfnamefont {R.}~\bibnamefont {Moessner}},\ }\href
  {\doibase 10.1103/PhysRevLett.115.030402} {\bibfield  {journal} {\bibinfo
  {journal} {Phys. Rev. Lett.}\ }\textbf {\bibinfo {volume} {115}},\ \bibinfo
  {pages} {030402} (\bibinfo {year} {2015})},\ \Eprint
  {http://arxiv.org/abs/1410.3455} {arXiv:1410.3455 [cond-mat.stat-mech]}
  \BibitemShut {NoStop}%
\bibitem [{\citenamefont {Ponte}\ \emph {et~al.}(2015)\citenamefont {Ponte},
  \citenamefont {Chandran}, \citenamefont {Papi\'{c}},\ and\ \citenamefont
  {Abanin}}]{Ponte2015}%
  \BibitemOpen
  \bibfield  {author} {\bibinfo {author} {\bibfnamefont {P.}~\bibnamefont
  {Ponte}}, \bibinfo {author} {\bibfnamefont {A.}~\bibnamefont {Chandran}},
  \bibinfo {author} {\bibfnamefont {Z.}~\bibnamefont {Papi\'{c}}}, \ and\
  \bibinfo {author} {\bibfnamefont {D.~A.}\ \bibnamefont {Abanin}},\ }\href
  {\doibase 10.1016/j.aop.2014.11.008} {\bibfield  {journal} {\bibinfo
  {journal} {Ann. Phys.}\ }\textbf {\bibinfo {volume} {353}},\ \bibinfo {pages}
  {196 } (\bibinfo {year} {2015})},\ \Eprint {http://arxiv.org/abs/1403.6480}
  {arXiv:1403.6480 [cond-mat.dis-nn]} \BibitemShut {NoStop}%
\bibitem [{\citenamefont {Bukov}\ \emph {et~al.}(2016)\citenamefont {Bukov},
  \citenamefont {Heyl}, \citenamefont {Huse},\ and\ \citenamefont
  {Polkovnikov}}]{Bukov2015c}%
  \BibitemOpen
  \bibfield  {author} {\bibinfo {author} {\bibfnamefont {M.}~\bibnamefont
  {Bukov}}, \bibinfo {author} {\bibfnamefont {M.}~\bibnamefont {Heyl}},
  \bibinfo {author} {\bibfnamefont {D.~A.}\ \bibnamefont {Huse}}, \ and\
  \bibinfo {author} {\bibfnamefont {A.}~\bibnamefont {Polkovnikov}},\ }\href
  {\doibase 10.1103/PhysRevB.93.155132} {\bibfield  {journal} {\bibinfo
  {journal} {Phys. Rev. B}\ }\textbf {\bibinfo {volume} {93}},\ \bibinfo
  {pages} {155132} (\bibinfo {year} {2016})},\ \Eprint
  {http://arxiv.org/abs/1512.02119} {arXiv:1512.02119 [cond-mat.quant-gas]}
  \BibitemShut {NoStop}%
\bibitem [{\citenamefont {{Clark}}\ \emph {et~al.}(2017)\citenamefont
  {{Clark}}, \citenamefont {{Gaj}}, \citenamefont {{Feng}},\ and\ \citenamefont
  {{Chin}}}]{Clark2017}%
  \BibitemOpen
  \bibfield  {author} {\bibinfo {author} {\bibfnamefont {L.~W.}\ \bibnamefont
  {{Clark}}}, \bibinfo {author} {\bibfnamefont {A.}~\bibnamefont {{Gaj}}},
  \bibinfo {author} {\bibfnamefont {L.}~\bibnamefont {{Feng}}}, \ and\ \bibinfo
  {author} {\bibfnamefont {C.}~\bibnamefont {{Chin}}},\ }\href {\doibase
  10.1038/nature24272} {\bibfield  {journal} {\bibinfo  {journal} {Nature}\
  }\textbf {\bibinfo {volume} {551}},\ \bibinfo {pages} {356} (\bibinfo {year}
  {2017})},\ \Eprint {http://arxiv.org/abs/1706.05560} {arXiv:1706.05560
  [cond-mat.quant-gas]} \BibitemShut {NoStop}%
\bibitem [{\citenamefont {Kandelaki}\ and\ \citenamefont
  {Rudner}(2018)}]{Kandelaki2017}%
  \BibitemOpen
  \bibfield  {author} {\bibinfo {author} {\bibfnamefont {E.}~\bibnamefont
  {Kandelaki}}\ and\ \bibinfo {author} {\bibfnamefont {M.~S.}\ \bibnamefont
  {Rudner}},\ }\href {\doibase 10.1103/PhysRevLett.121.036801} {\bibfield
  {journal} {\bibinfo  {journal} {Phys. Rev. Lett.}\ }\textbf {\bibinfo
  {volume} {121}},\ \bibinfo {pages} {036801} (\bibinfo {year} {2018})},\
  \Eprint {http://arxiv.org/abs/1709.04448} {arXiv:1709.04448
  [cond-mat.mes-hall]} \BibitemShut {NoStop}%
\bibitem [{\citenamefont {Shibata}\ \emph {et~al.}(2019)\citenamefont
  {Shibata}, \citenamefont {Torii}, \citenamefont {Shibayama}, \citenamefont
  {Eto}, \citenamefont {Saito},\ and\ \citenamefont {Hirano}}]{Shibata2018}%
  \BibitemOpen
  \bibfield  {author} {\bibinfo {author} {\bibfnamefont {K.}~\bibnamefont
  {Shibata}}, \bibinfo {author} {\bibfnamefont {A.}~\bibnamefont {Torii}},
  \bibinfo {author} {\bibfnamefont {H.}~\bibnamefont {Shibayama}}, \bibinfo
  {author} {\bibfnamefont {Y.}~\bibnamefont {Eto}}, \bibinfo {author}
  {\bibfnamefont {H.}~\bibnamefont {Saito}}, \ and\ \bibinfo {author}
  {\bibfnamefont {T.}~\bibnamefont {Hirano}},\ }\href {\doibase
  10.1103/PhysRevA.99.013622} {\bibfield  {journal} {\bibinfo  {journal} {Phys.
  Rev. A}\ }\textbf {\bibinfo {volume} {99}},\ \bibinfo {pages} {013622}
  (\bibinfo {year} {2019})},\ \Eprint {http://arxiv.org/abs/1812.01284}
  {arXiv:1812.01284 [physics.atom-ph]} \BibitemShut {NoStop}%
\bibitem [{\citenamefont {Dehghani}\ \emph {et~al.}(2014)\citenamefont
  {Dehghani}, \citenamefont {Oka},\ and\ \citenamefont {Mitra}}]{Dehghani2014}%
  \BibitemOpen
  \bibfield  {author} {\bibinfo {author} {\bibfnamefont {H.}~\bibnamefont
  {Dehghani}}, \bibinfo {author} {\bibfnamefont {T.}~\bibnamefont {Oka}}, \
  and\ \bibinfo {author} {\bibfnamefont {A.}~\bibnamefont {Mitra}},\ }\href
  {\doibase 10.1103/PhysRevB.90.195429} {\bibfield  {journal} {\bibinfo
  {journal} {Phys. Rev. B}\ }\textbf {\bibinfo {volume} {90}},\ \bibinfo
  {pages} {195429} (\bibinfo {year} {2014})},\ \Eprint
  {http://arxiv.org/abs/1406.6626} {arXiv:1406.6626 [cond-mat.mes-hall]}
  \BibitemShut {NoStop}%
\bibitem [{\citenamefont {Knap}\ \emph {et~al.}(2016)\citenamefont {Knap},
  \citenamefont {Babadi}, \citenamefont {Refael}, \citenamefont {Martin},\ and\
  \citenamefont {Demler}}]{Knap2016}%
  \BibitemOpen
  \bibfield  {author} {\bibinfo {author} {\bibfnamefont {M.}~\bibnamefont
  {Knap}}, \bibinfo {author} {\bibfnamefont {M.}~\bibnamefont {Babadi}},
  \bibinfo {author} {\bibfnamefont {G.}~\bibnamefont {Refael}}, \bibinfo
  {author} {\bibfnamefont {I.}~\bibnamefont {Martin}}, \ and\ \bibinfo {author}
  {\bibfnamefont {E.}~\bibnamefont {Demler}},\ }\href {\doibase
  10.1103/PhysRevB.94.214504} {\bibfield  {journal} {\bibinfo  {journal} {Phys.
  Rev. B}\ }\textbf {\bibinfo {volume} {94}},\ \bibinfo {pages} {214504}
  (\bibinfo {year} {2016})},\ \Eprint {http://arxiv.org/abs/1511.07874}
  {arXiv:1511.07874 [cond-mat.supr-con]} \BibitemShut {NoStop}%
\bibitem [{\citenamefont {Babadi}\ \emph {et~al.}(2017)\citenamefont {Babadi},
  \citenamefont {Knap}, \citenamefont {Martin}, \citenamefont {Refael},\ and\
  \citenamefont {Demler}}]{Babadi2017}%
  \BibitemOpen
  \bibfield  {author} {\bibinfo {author} {\bibfnamefont {M.}~\bibnamefont
  {Babadi}}, \bibinfo {author} {\bibfnamefont {M.}~\bibnamefont {Knap}},
  \bibinfo {author} {\bibfnamefont {I.}~\bibnamefont {Martin}}, \bibinfo
  {author} {\bibfnamefont {G.}~\bibnamefont {Refael}}, \ and\ \bibinfo {author}
  {\bibfnamefont {E.}~\bibnamefont {Demler}},\ }\href {\doibase
  10.1103/PhysRevB.96.014512} {\bibfield  {journal} {\bibinfo  {journal} {Phys.
  Rev. B}\ }\textbf {\bibinfo {volume} {96}},\ \bibinfo {pages} {014512}
  (\bibinfo {year} {2017})},\ \Eprint {http://arxiv.org/abs/1702.02531}
  {arXiv:1702.02531 [cond-mat.supr-con]} \BibitemShut {NoStop}%
\bibitem [{\citenamefont {Murakami}\ \emph {et~al.}(2017)\citenamefont
  {Murakami}, \citenamefont {Tsuji}, \citenamefont {Eckstein},\ and\
  \citenamefont {Werner}}]{Murakami2017}%
  \BibitemOpen
  \bibfield  {author} {\bibinfo {author} {\bibfnamefont {Y.}~\bibnamefont
  {Murakami}}, \bibinfo {author} {\bibfnamefont {N.}~\bibnamefont {Tsuji}},
  \bibinfo {author} {\bibfnamefont {M.}~\bibnamefont {Eckstein}}, \ and\
  \bibinfo {author} {\bibfnamefont {P.}~\bibnamefont {Werner}},\ }\href
  {\doibase 10.1103/PhysRevB.96.045125} {\bibfield  {journal} {\bibinfo
  {journal} {Phys. Rev. B}\ }\textbf {\bibinfo {volume} {96}},\ \bibinfo
  {pages} {045125} (\bibinfo {year} {2017})},\ \Eprint
  {http://arxiv.org/abs/1702.02942} {arXiv:1702.02942 [cond-mat.supr-con]}
  \BibitemShut {NoStop}%
\bibitem [{\citenamefont {Seetharam}\ \emph {et~al.}(2019)\citenamefont
  {Seetharam}, \citenamefont {Bardyn}, \citenamefont {Lindner}, \citenamefont
  {Rudner},\ and\ \citenamefont {Refael}}]{Seetharam2018}%
  \BibitemOpen
  \bibfield  {author} {\bibinfo {author} {\bibfnamefont {K.~I.}\ \bibnamefont
  {Seetharam}}, \bibinfo {author} {\bibfnamefont {C.-E.}\ \bibnamefont
  {Bardyn}}, \bibinfo {author} {\bibfnamefont {N.~H.}\ \bibnamefont {Lindner}},
  \bibinfo {author} {\bibfnamefont {M.~S.}\ \bibnamefont {Rudner}}, \ and\
  \bibinfo {author} {\bibfnamefont {G.}~\bibnamefont {Refael}},\ }\href
  {\doibase 10.1103/PhysRevB.99.014307} {\bibfield  {journal} {\bibinfo
  {journal} {Phys. Rev. B}\ }\textbf {\bibinfo {volume} {99}},\ \bibinfo
  {pages} {014307} (\bibinfo {year} {2019})},\ \Eprint
  {http://arxiv.org/abs/1806.10620} {arXiv:1806.10620 [cond-mat.mes-hall]}
  \BibitemShut {NoStop}%
\bibitem [{\citenamefont {Xu}\ \emph {et~al.}(2015)\citenamefont {Xu},
  \citenamefont {Gullans},\ and\ \citenamefont {Taylor}}]{Xu2015}%
  \BibitemOpen
  \bibfield  {author} {\bibinfo {author} {\bibfnamefont {X.}~\bibnamefont
  {Xu}}, \bibinfo {author} {\bibfnamefont {M.}~\bibnamefont {Gullans}}, \ and\
  \bibinfo {author} {\bibfnamefont {J.~M.}\ \bibnamefont {Taylor}},\ }\href
  {\doibase 10.1103/PhysRevA.91.013818} {\bibfield  {journal} {\bibinfo
  {journal} {Phys. Rev. A}\ }\textbf {\bibinfo {volume} {91}},\ \bibinfo
  {pages} {013818} (\bibinfo {year} {2015})},\ \Eprint
  {http://arxiv.org/abs/1404.3726} {arXiv:1404.3726 [quant-ph]} \BibitemShut
  {NoStop}%
\bibitem [{\citenamefont {Chitra}\ and\ \citenamefont
  {Zilberberg}(2015)}]{Chitra2015}%
  \BibitemOpen
  \bibfield  {author} {\bibinfo {author} {\bibfnamefont {R.}~\bibnamefont
  {Chitra}}\ and\ \bibinfo {author} {\bibfnamefont {O.}~\bibnamefont
  {Zilberberg}},\ }\href {\doibase 10.1103/PhysRevA.92.023815} {\bibfield
  {journal} {\bibinfo  {journal} {Phys. Rev. A}\ }\textbf {\bibinfo {volume}
  {92}},\ \bibinfo {pages} {023815} (\bibinfo {year} {2015})},\ \Eprint
  {http://arxiv.org/abs/1501.07098} {arXiv:1501.07098 [cond-mat.quant-gas]}
  \BibitemShut {NoStop}%
\bibitem [{\citenamefont {{Lemonde}}\ \emph {et~al.}(2016)\citenamefont
  {{Lemonde}}, \citenamefont {{Didier}},\ and\ \citenamefont
  {{Clerk}}}]{Lemonde2016}%
  \BibitemOpen
  \bibfield  {author} {\bibinfo {author} {\bibfnamefont {M.-A.}\ \bibnamefont
  {{Lemonde}}}, \bibinfo {author} {\bibfnamefont {N.}~\bibnamefont {{Didier}}},
  \ and\ \bibinfo {author} {\bibfnamefont {A.~A.}\ \bibnamefont {{Clerk}}},\
  }\href {\doibase 10.1038/ncomms11338} {\bibfield  {journal} {\bibinfo
  {journal} {Nat. Commun.}\ }\textbf {\bibinfo {volume} {7}},\ \bibinfo {eid}
  {11338} (\bibinfo {year} {2016})},\ \Eprint {http://arxiv.org/abs/1509.09238}
  {arXiv:1509.09238 [quant-ph]} \BibitemShut {NoStop}%
\bibitem [{\citenamefont {Stehlik}\ \emph {et~al.}(2016)\citenamefont
  {Stehlik}, \citenamefont {Liu}, \citenamefont {Eichler}, \citenamefont
  {Hartke}, \citenamefont {Mi}, \citenamefont {Gullans}, \citenamefont
  {Taylor},\ and\ \citenamefont {Petta}}]{Stehlik2016}%
  \BibitemOpen
  \bibfield  {author} {\bibinfo {author} {\bibfnamefont {J.}~\bibnamefont
  {Stehlik}}, \bibinfo {author} {\bibfnamefont {Y.-Y.}\ \bibnamefont {Liu}},
  \bibinfo {author} {\bibfnamefont {C.}~\bibnamefont {Eichler}}, \bibinfo
  {author} {\bibfnamefont {T.~R.}\ \bibnamefont {Hartke}}, \bibinfo {author}
  {\bibfnamefont {X.}~\bibnamefont {Mi}}, \bibinfo {author} {\bibfnamefont
  {M.~J.}\ \bibnamefont {Gullans}}, \bibinfo {author} {\bibfnamefont {J.~M.}\
  \bibnamefont {Taylor}}, \ and\ \bibinfo {author} {\bibfnamefont {J.~R.}\
  \bibnamefont {Petta}},\ }\href {\doibase 10.1103/PhysRevX.6.041027}
  {\bibfield  {journal} {\bibinfo  {journal} {Phys. Rev. X}\ }\textbf {\bibinfo
  {volume} {6}},\ \bibinfo {pages} {041027} (\bibinfo {year} {2016})},\ \Eprint
  {http://arxiv.org/abs/1607.08229} {arXiv:1607.08229 [cond-mat.mes-hall]}
  \BibitemShut {NoStop}%
\bibitem [{\citenamefont {Gong}\ \emph {et~al.}(2018)\citenamefont {Gong},
  \citenamefont {Hamazaki},\ and\ \citenamefont {Ueda}}]{Gong2018}%
  \BibitemOpen
  \bibfield  {author} {\bibinfo {author} {\bibfnamefont {Z.}~\bibnamefont
  {Gong}}, \bibinfo {author} {\bibfnamefont {R.}~\bibnamefont {Hamazaki}}, \
  and\ \bibinfo {author} {\bibfnamefont {M.}~\bibnamefont {Ueda}},\ }\href
  {\doibase 10.1103/PhysRevLett.120.040404} {\bibfield  {journal} {\bibinfo
  {journal} {Phys. Rev. Lett.}\ }\textbf {\bibinfo {volume} {120}},\ \bibinfo
  {pages} {040404} (\bibinfo {year} {2018})},\ \Eprint
  {http://arxiv.org/abs/1708.01472} {arXiv:1708.01472 [cond-mat.stat-mech]}
  \BibitemShut {NoStop}%
\bibitem [{\citenamefont {H{\"a}nggi}\ and\ \citenamefont
  {Marchesoni}(2009)}]{Hanggi2009}%
  \BibitemOpen
  \bibfield  {author} {\bibinfo {author} {\bibfnamefont {P.}~\bibnamefont
  {H{\"a}nggi}}\ and\ \bibinfo {author} {\bibfnamefont {F.}~\bibnamefont
  {Marchesoni}},\ }\href {\doibase 10.1103/RevModPhys.81.387} {\bibfield
  {journal} {\bibinfo  {journal} {Rev. Mod. Phys.}\ }\textbf {\bibinfo {volume}
  {81}},\ \bibinfo {pages} {387} (\bibinfo {year} {2009})},\ \Eprint
  {http://arxiv.org/abs/0807.1283} {arXiv:0807.1283 [cond-mat.stat-mech]}
  \BibitemShut {NoStop}%
\bibitem [{\citenamefont {Salger}\ \emph {et~al.}(2013)\citenamefont {Salger},
  \citenamefont {Kling}, \citenamefont {Denisov}, \citenamefont {Ponomarev},
  \citenamefont {H\"anggi},\ and\ \citenamefont {Weitz}}]{Salger2013}%
  \BibitemOpen
  \bibfield  {author} {\bibinfo {author} {\bibfnamefont {T.}~\bibnamefont
  {Salger}}, \bibinfo {author} {\bibfnamefont {S.}~\bibnamefont {Kling}},
  \bibinfo {author} {\bibfnamefont {S.}~\bibnamefont {Denisov}}, \bibinfo
  {author} {\bibfnamefont {A.~V.}\ \bibnamefont {Ponomarev}}, \bibinfo {author}
  {\bibfnamefont {P.}~\bibnamefont {H\"anggi}}, \ and\ \bibinfo {author}
  {\bibfnamefont {M.}~\bibnamefont {Weitz}},\ }\href {\doibase
  10.1103/PhysRevLett.110.135302} {\bibfield  {journal} {\bibinfo  {journal}
  {Phys. Rev. Lett.}\ }\textbf {\bibinfo {volume} {110}},\ \bibinfo {pages}
  {135302} (\bibinfo {year} {2013})},\ \Eprint {http://arxiv.org/abs/1202.5174}
  {arXiv:1202.5174 [cond-mat.quant-gas]} \BibitemShut {NoStop}%
\bibitem [{\citenamefont {Denisov}\ \emph {et~al.}(2014)\citenamefont
  {Denisov}, \citenamefont {Flach},\ and\ \citenamefont
  {H{\"a}nggi}}]{Denisov2014}%
  \BibitemOpen
  \bibfield  {author} {\bibinfo {author} {\bibfnamefont {S.}~\bibnamefont
  {Denisov}}, \bibinfo {author} {\bibfnamefont {S.}~\bibnamefont {Flach}}, \
  and\ \bibinfo {author} {\bibfnamefont {P.}~\bibnamefont {H{\"a}nggi}},\
  }\href {\doibase 10.1016/j.physrep.2014.01.003} {\bibfield  {journal}
  {\bibinfo  {journal} {Phys. Rep.}\ }\textbf {\bibinfo {volume} {538}},\
  \bibinfo {pages} {77 } (\bibinfo {year} {2014})},\ \Eprint
  {http://arxiv.org/abs/1311.1086} {arXiv:1311.1086 [cond-mat.mes-hall]}
  \BibitemShut {NoStop}%
\bibitem [{\citenamefont {Seetharam}\ \emph {et~al.}(2015)\citenamefont
  {Seetharam}, \citenamefont {Bardyn}, \citenamefont {Lindner}, \citenamefont
  {Rudner},\ and\ \citenamefont {Refael}}]{Seetharam2015}%
  \BibitemOpen
  \bibfield  {author} {\bibinfo {author} {\bibfnamefont {K.~I.}\ \bibnamefont
  {Seetharam}}, \bibinfo {author} {\bibfnamefont {C.-E.}\ \bibnamefont
  {Bardyn}}, \bibinfo {author} {\bibfnamefont {N.~H.}\ \bibnamefont {Lindner}},
  \bibinfo {author} {\bibfnamefont {M.~S.}\ \bibnamefont {Rudner}}, \ and\
  \bibinfo {author} {\bibfnamefont {G.}~\bibnamefont {Refael}},\ }\href
  {\doibase 10.1103/PhysRevX.5.041050} {\bibfield  {journal} {\bibinfo
  {journal} {Phys. Rev. X}\ }\textbf {\bibinfo {volume} {5}},\ \bibinfo {pages}
  {041050} (\bibinfo {year} {2015})},\ \Eprint
  {http://arxiv.org/abs/1502.02664} {arXiv:1502.02664 [cond-mat.mes-hall]}
  \BibitemShut {NoStop}%
\bibitem [{\citenamefont {Lazarides}\ and\ \citenamefont
  {Moessner}(2017)}]{Lazarides2017}%
  \BibitemOpen
  \bibfield  {author} {\bibinfo {author} {\bibfnamefont {A.}~\bibnamefont
  {Lazarides}}\ and\ \bibinfo {author} {\bibfnamefont {R.}~\bibnamefont
  {Moessner}},\ }\href {\doibase 10.1103/PhysRevB.95.195135} {\bibfield
  {journal} {\bibinfo  {journal} {Phys. Rev. B}\ }\textbf {\bibinfo {volume}
  {95}},\ \bibinfo {pages} {195135} (\bibinfo {year} {2017})},\ \Eprint
  {http://arxiv.org/abs/1703.02547} {arXiv:1703.02547 [cond-mat.stat-mech]}
  \BibitemShut {NoStop}%
\bibitem [{\citenamefont {{Lerose}}\ \emph {et~al.}()\citenamefont {{Lerose}},
  \citenamefont {{Marino}}, \citenamefont {{Gambassi}},\ and\ \citenamefont
  {{Silva}}}]{Lerose2018}%
  \BibitemOpen
  \bibfield  {author} {\bibinfo {author} {\bibfnamefont {A.}~\bibnamefont
  {{Lerose}}}, \bibinfo {author} {\bibfnamefont {J.}~\bibnamefont {{Marino}}},
  \bibinfo {author} {\bibfnamefont {A.}~\bibnamefont {{Gambassi}}}, \ and\
  \bibinfo {author} {\bibfnamefont {A.}~\bibnamefont {{Silva}}},\ }\href@noop
  {} {\bibfield  {journal} {\bibinfo  {journal} {arXiv:1803.04490
  [cond-mat.stat-mech]}\ }}\Eprint {http://arxiv.org/abs/1803.04490}
  {arXiv:1803.04490 [cond-mat.stat-mech]} \BibitemShut {NoStop}%
\bibitem [{\citenamefont {Li}\ \emph {et~al.}(2019)\citenamefont {Li},
  \citenamefont {Harter}, \citenamefont {Liu}, \citenamefont {de~Melo},
  \citenamefont {Joglekar},\ and\ \citenamefont {Luo}}]{Li2016}%
  \BibitemOpen
  \bibfield  {author} {\bibinfo {author} {\bibfnamefont {J.}~\bibnamefont
  {Li}}, \bibinfo {author} {\bibfnamefont {A.~K.}\ \bibnamefont {Harter}},
  \bibinfo {author} {\bibfnamefont {J.}~\bibnamefont {Liu}}, \bibinfo {author}
  {\bibfnamefont {L.}~\bibnamefont {de~Melo}}, \bibinfo {author} {\bibfnamefont
  {Y.~N.}\ \bibnamefont {Joglekar}}, \ and\ \bibinfo {author} {\bibfnamefont
  {L.}~\bibnamefont {Luo}},\ }\href {\doibase 10.1038/s41467-019-08596-1}
  {\bibfield  {journal} {\bibinfo  {journal} {Nature Commun.}\ }\textbf
  {\bibinfo {volume} {10}},\ \bibinfo {pages} {855} (\bibinfo {year} {2019})},\
  \Eprint {http://arxiv.org/abs/1608.05061} {arXiv:1608.05061
  [cond-mat.quant-gas]} \BibitemShut {NoStop}%
\bibitem [{\citenamefont {Iwahori}\ and\ \citenamefont
  {Kawakami}(2017)}]{Iwahori2017}%
  \BibitemOpen
  \bibfield  {author} {\bibinfo {author} {\bibfnamefont {K.}~\bibnamefont
  {Iwahori}}\ and\ \bibinfo {author} {\bibfnamefont {N.}~\bibnamefont
  {Kawakami}},\ }\href {\doibase 10.1103/PhysRevA.95.043621} {\bibfield
  {journal} {\bibinfo  {journal} {Phys. Rev. A}\ }\textbf {\bibinfo {volume}
  {95}},\ \bibinfo {pages} {043621} (\bibinfo {year} {2017})},\ \Eprint
  {http://arxiv.org/abs/1702.03506} {arXiv:1702.03506 [cond-mat.stat-mech]}
  \BibitemShut {NoStop}%
\bibitem [{\citenamefont {{Tomita}}\ \emph {et~al.}(2017)\citenamefont
  {{Tomita}}, \citenamefont {{Nakajima}}, \citenamefont {{Danshita}},
  \citenamefont {{Takasu}},\ and\ \citenamefont {{Takahashi}}}]{Tomita2017}%
  \BibitemOpen
  \bibfield  {author} {\bibinfo {author} {\bibfnamefont {T.}~\bibnamefont
  {{Tomita}}}, \bibinfo {author} {\bibfnamefont {S.}~\bibnamefont
  {{Nakajima}}}, \bibinfo {author} {\bibfnamefont {I.}~\bibnamefont
  {{Danshita}}}, \bibinfo {author} {\bibfnamefont {Y.}~\bibnamefont
  {{Takasu}}}, \ and\ \bibinfo {author} {\bibfnamefont {Y.}~\bibnamefont
  {{Takahashi}}},\ }\href {\doibase 10.1126/sciadv.1701513} {\bibfield
  {journal} {\bibinfo  {journal} {Sci. Adv.}\ }\textbf {\bibinfo {volume}
  {3}},\ \bibinfo {pages} {e1701513} (\bibinfo {year} {2017})},\ \Eprint
  {http://arxiv.org/abs/1705.09942} {arXiv:1705.09942 [cond-mat.quant-gas]}
  \BibitemShut {NoStop}%
\bibitem [{\citenamefont {{De Sarkar}}\ \emph {et~al.}(2014)\citenamefont {{De
  Sarkar}}, \citenamefont {Sensarma},\ and\ \citenamefont
  {Sengupta}}]{DeSarkar2014}%
  \BibitemOpen
  \bibfield  {author} {\bibinfo {author} {\bibfnamefont {S.}~\bibnamefont {{De
  Sarkar}}}, \bibinfo {author} {\bibfnamefont {R.}~\bibnamefont {Sensarma}}, \
  and\ \bibinfo {author} {\bibfnamefont {K.}~\bibnamefont {Sengupta}},\ }\href
  {\doibase 10.1088/0953-8984/26/32/325602} {\bibfield  {journal} {\bibinfo
  {journal} {J. Phys. Condens. Matter}\ }\textbf {\bibinfo {volume} {26}},\
  \bibinfo {pages} {325602} (\bibinfo {year} {2014})},\ \Eprint
  {http://arxiv.org/abs/1308.4689} {arXiv:1308.4689 [cond-mat.str-el]}
  \BibitemShut {NoStop}%
\bibitem [{\citenamefont {Nikoghosyan}\ \emph {et~al.}(2016)\citenamefont
  {Nikoghosyan}, \citenamefont {Nigmatullin},\ and\ \citenamefont
  {Plenio}}]{Nikoghosyan2016}%
  \BibitemOpen
  \bibfield  {author} {\bibinfo {author} {\bibfnamefont {G.}~\bibnamefont
  {Nikoghosyan}}, \bibinfo {author} {\bibfnamefont {R.}~\bibnamefont
  {Nigmatullin}}, \ and\ \bibinfo {author} {\bibfnamefont {M.~B.}\ \bibnamefont
  {Plenio}},\ }\href {\doibase 10.1103/PhysRevLett.116.080601} {\bibfield
  {journal} {\bibinfo  {journal} {Phys. Rev. Lett.}\ }\textbf {\bibinfo
  {volume} {116}},\ \bibinfo {pages} {080601} (\bibinfo {year} {2016})},\
  \Eprint {http://arxiv.org/abs/1311.1543} {arXiv:1311.1543
  [cond-mat.stat-mech]} \BibitemShut {NoStop}%
\bibitem [{\citenamefont {Feng}\ \emph {et~al.}(2016)\citenamefont {Feng},
  \citenamefont {Yin},\ and\ \citenamefont {Zhong}}]{Baoquan2016}%
  \BibitemOpen
  \bibfield  {author} {\bibinfo {author} {\bibfnamefont {B.}~\bibnamefont
  {Feng}}, \bibinfo {author} {\bibfnamefont {S.}~\bibnamefont {Yin}}, \ and\
  \bibinfo {author} {\bibfnamefont {F.}~\bibnamefont {Zhong}},\ }\href
  {\doibase 10.1103/PhysRevB.94.144103} {\bibfield  {journal} {\bibinfo
  {journal} {Phys. Rev. B}\ }\textbf {\bibinfo {volume} {94}},\ \bibinfo
  {pages} {144103} (\bibinfo {year} {2016})},\ \Eprint
  {http://arxiv.org/abs/1604.04345} {arXiv:1604.04345 [cond-mat.stat-mech]}
  \BibitemShut {NoStop}%
\bibitem [{\citenamefont {Korniss}\ \emph {et~al.}(2000)\citenamefont
  {Korniss}, \citenamefont {White}, \citenamefont {Rikvold},\ and\
  \citenamefont {Novotny}}]{Korniss2000}%
  \BibitemOpen
  \bibfield  {author} {\bibinfo {author} {\bibfnamefont {G.}~\bibnamefont
  {Korniss}}, \bibinfo {author} {\bibfnamefont {C.~J.}\ \bibnamefont {White}},
  \bibinfo {author} {\bibfnamefont {P.~A.}\ \bibnamefont {Rikvold}}, \ and\
  \bibinfo {author} {\bibfnamefont {M.~A.}\ \bibnamefont {Novotny}},\ }\href
  {\doibase 10.1103/PhysRevE.63.016120} {\bibfield  {journal} {\bibinfo
  {journal} {Phys. Rev. E}\ }\textbf {\bibinfo {volume} {63}},\ \bibinfo
  {pages} {016120} (\bibinfo {year} {2000})},\ \Eprint
  {http://arxiv.org/abs/cond-mat/0008155} {cond-mat/0008155} \BibitemShut
  {NoStop}%
\bibitem [{\citenamefont {Fujisaka}\ \emph {et~al.}(2001)\citenamefont
  {Fujisaka}, \citenamefont {Tutu},\ and\ \citenamefont
  {Rikvold}}]{Fujisaka2001}%
  \BibitemOpen
  \bibfield  {author} {\bibinfo {author} {\bibfnamefont {H.}~\bibnamefont
  {Fujisaka}}, \bibinfo {author} {\bibfnamefont {H.}~\bibnamefont {Tutu}}, \
  and\ \bibinfo {author} {\bibfnamefont {P.~A.}\ \bibnamefont {Rikvold}},\
  }\href {\doibase 10.1103/PhysRevE.63.036109} {\bibfield  {journal} {\bibinfo
  {journal} {Phys. Rev. E}\ }\textbf {\bibinfo {volume} {63}},\ \bibinfo
  {pages} {036109} (\bibinfo {year} {2001})},\ \Eprint
  {http://arxiv.org/abs/cond-mat/0009284} {cond-mat/0009284} \BibitemShut
  {NoStop}%
\bibitem [{\citenamefont {Buend\'{\i}a}\ and\ \citenamefont
  {Rikvold}(2008)}]{Buendia2008}%
  \BibitemOpen
  \bibfield  {author} {\bibinfo {author} {\bibfnamefont {G.~M.}\ \bibnamefont
  {Buend\'{\i}a}}\ and\ \bibinfo {author} {\bibfnamefont {P.~A.}\ \bibnamefont
  {Rikvold}},\ }\href {\doibase 10.1103/PhysRevE.78.051108} {\bibfield
  {journal} {\bibinfo  {journal} {Phys. Rev. E}\ }\textbf {\bibinfo {volume}
  {78}},\ \bibinfo {pages} {051108} (\bibinfo {year} {2008})},\ \Eprint
  {http://arxiv.org/abs/0809.0523} {arXiv:0809.0523 [cond-mat.stat-mech]}
  \BibitemShut {NoStop}%
\bibitem [{foo({\natexlab{a}})}]{footnote4}%
  \BibitemOpen
  \href@noop {} {} \bibinfo {note} {The applicability of the
  $\epsilon$ expansion is a property of the \WF fixed point. This expansion
  works because the fixed-point coupling vanishes as $d \rightarrow 4$, \ie
  ${g^* \sim \epsilon}$. Here we investigate near-equilibrium critical physics
  in the sense that the relevant \RG fixed point is the same as at equilibrium.
  Therefore the $\epsilon$ expansion remains valid here.}\BibitemShut {Stop}%
\bibitem [{\citenamefont {Coleman}\ and\ \citenamefont
  {Weinberg}(1973)}]{Coleman1973}%
  \BibitemOpen
  \bibfield  {author} {\bibinfo {author} {\bibfnamefont {S.}~\bibnamefont
  {Coleman}}\ and\ \bibinfo {author} {\bibfnamefont {E.}~\bibnamefont
  {Weinberg}},\ }\href {\doibase 10.1103/PhysRevD.7.1888} {\bibfield  {journal}
  {\bibinfo  {journal} {Phys. Rev. D}\ }\textbf {\bibinfo {volume} {7}},\
  \bibinfo {pages} {1888} (\bibinfo {year} {1973})}\BibitemShut {NoStop}%
\bibitem [{\citenamefont {Halperin}\ \emph {et~al.}(1974)\citenamefont
  {Halperin}, \citenamefont {Lubensky},\ and\ \citenamefont
  {Ma}}]{Halperin1974}%
  \BibitemOpen
  \bibfield  {author} {\bibinfo {author} {\bibfnamefont {B.~I.}\ \bibnamefont
  {Halperin}}, \bibinfo {author} {\bibfnamefont {T.~C.}\ \bibnamefont
  {Lubensky}}, \ and\ \bibinfo {author} {\bibfnamefont {S.-K.}\ \bibnamefont
  {Ma}},\ }\href {\doibase 10.1103/PhysRevLett.32.292} {\bibfield  {journal}
  {\bibinfo  {journal} {Phys. Rev. Lett.}\ }\textbf {\bibinfo {volume} {32}},\
  \bibinfo {pages} {292} (\bibinfo {year} {1974})}\BibitemShut {NoStop}%
\bibitem [{\citenamefont {Fisher}\ and\ \citenamefont
  {Nelson}(1974)}]{Fisher1974}%
  \BibitemOpen
  \bibfield  {author} {\bibinfo {author} {\bibfnamefont {M.~E.}\ \bibnamefont
  {Fisher}}\ and\ \bibinfo {author} {\bibfnamefont {D.~R.}\ \bibnamefont
  {Nelson}},\ }\href {\doibase 10.1103/PhysRevLett.32.1350} {\bibfield
  {journal} {\bibinfo  {journal} {Phys. Rev. Lett.}\ }\textbf {\bibinfo
  {volume} {32}},\ \bibinfo {pages} {1350} (\bibinfo {year}
  {1974})}\BibitemShut {NoStop}%
\bibitem [{\citenamefont {Nelson}\ \emph {et~al.}(1974)\citenamefont {Nelson},
  \citenamefont {Kosterlitz},\ and\ \citenamefont {Fisher}}]{Nelson1974}%
  \BibitemOpen
  \bibfield  {author} {\bibinfo {author} {\bibfnamefont {D.~R.}\ \bibnamefont
  {Nelson}}, \bibinfo {author} {\bibfnamefont {J.~M.}\ \bibnamefont
  {Kosterlitz}}, \ and\ \bibinfo {author} {\bibfnamefont {M.~E.}\ \bibnamefont
  {Fisher}},\ }\href {\doibase 10.1103/PhysRevLett.33.813} {\bibfield
  {journal} {\bibinfo  {journal} {Phys. Rev. Lett.}\ }\textbf {\bibinfo
  {volume} {33}},\ \bibinfo {pages} {813} (\bibinfo {year} {1974})}\BibitemShut
  {NoStop}%
\bibitem [{\citenamefont {Golner}(1973)}]{Golner1973}%
  \BibitemOpen
  \bibfield  {author} {\bibinfo {author} {\bibfnamefont {G.~R.}\ \bibnamefont
  {Golner}},\ }\href {\doibase 10.1103/PhysRevB.8.3419} {\bibfield  {journal}
  {\bibinfo  {journal} {Phys. Rev. B}\ }\textbf {\bibinfo {volume} {8}},\
  \bibinfo {pages} {3419} (\bibinfo {year} {1973})}\BibitemShut {NoStop}%
\bibitem [{\citenamefont {Wu}(1982)}]{Wu1982}%
  \BibitemOpen
  \bibfield  {author} {\bibinfo {author} {\bibfnamefont {F.~Y.}\ \bibnamefont
  {Wu}},\ }\href {\doibase 10.1103/RevModPhys.54.235} {\bibfield  {journal}
  {\bibinfo  {journal} {Rev. Mod. Phys.}\ }\textbf {\bibinfo {volume} {54}},\
  \bibinfo {pages} {235} (\bibinfo {year} {1982})}\BibitemShut {NoStop}%
\bibitem [{\citenamefont {Carmona}\ \emph {et~al.}(2000)\citenamefont
  {Carmona}, \citenamefont {Pelissetto},\ and\ \citenamefont
  {Vicari}}]{Carmona2000}%
  \BibitemOpen
  \bibfield  {author} {\bibinfo {author} {\bibfnamefont {J.~M.}\ \bibnamefont
  {Carmona}}, \bibinfo {author} {\bibfnamefont {A.}~\bibnamefont {Pelissetto}},
  \ and\ \bibinfo {author} {\bibfnamefont {E.}~\bibnamefont {Vicari}},\ }\href
  {\doibase 10.1103/PhysRevB.61.15136} {\bibfield  {journal} {\bibinfo
  {journal} {Phys. Rev. B}\ }\textbf {\bibinfo {volume} {61}},\ \bibinfo
  {pages} {15136} (\bibinfo {year} {2000})},\ \Eprint
  {http://arxiv.org/abs/cond-mat/9912115} {cond-mat/9912115} \BibitemShut
  {NoStop}%
\bibitem [{\citenamefont {Aharony}(2003)}]{Aharony2003}%
  \BibitemOpen
  \bibfield  {author} {\bibinfo {author} {\bibfnamefont {A.}~\bibnamefont
  {Aharony}},\ }\href {\doibase 10.1023/A:1022103717585} {\bibfield  {journal}
  {\bibinfo  {journal} {J. Stat. Phys.}\ }\textbf {\bibinfo {volume} {110}},\
  \bibinfo {pages} {659} (\bibinfo {year} {2003})},\ \Eprint
  {http://arxiv.org/abs/cond-mat/0201576} {cond-mat/0201576} \BibitemShut
  {NoStop}%
\bibitem [{\citenamefont {Vorberg}\ \emph {et~al.}(2013)\citenamefont
  {Vorberg}, \citenamefont {Wustmann}, \citenamefont {Ketzmerick},\ and\
  \citenamefont {Eckardt}}]{Vorberg2013}%
  \BibitemOpen
  \bibfield  {author} {\bibinfo {author} {\bibfnamefont {D.}~\bibnamefont
  {Vorberg}}, \bibinfo {author} {\bibfnamefont {W.}~\bibnamefont {Wustmann}},
  \bibinfo {author} {\bibfnamefont {R.}~\bibnamefont {Ketzmerick}}, \ and\
  \bibinfo {author} {\bibfnamefont {A.}~\bibnamefont {Eckardt}},\ }\href
  {\doibase 10.1103/PhysRevLett.111.240405} {\bibfield  {journal} {\bibinfo
  {journal} {Phys. Rev. Lett.}\ }\textbf {\bibinfo {volume} {111}},\ \bibinfo
  {pages} {240405} (\bibinfo {year} {2013})},\ \Eprint
  {http://arxiv.org/abs/1308.2776} {arXiv:1308.2776 [cond-mat.stat-mech]}
  \BibitemShut {NoStop}%
\bibitem [{\citenamefont {Vorberg}\ \emph {et~al.}(2015)\citenamefont
  {Vorberg}, \citenamefont {Wustmann}, \citenamefont {Schomerus}, \citenamefont
  {Ketzmerick},\ and\ \citenamefont {Eckardt}}]{Vorberg2015}%
  \BibitemOpen
  \bibfield  {author} {\bibinfo {author} {\bibfnamefont {D.}~\bibnamefont
  {Vorberg}}, \bibinfo {author} {\bibfnamefont {W.}~\bibnamefont {Wustmann}},
  \bibinfo {author} {\bibfnamefont {H.}~\bibnamefont {Schomerus}}, \bibinfo
  {author} {\bibfnamefont {R.}~\bibnamefont {Ketzmerick}}, \ and\ \bibinfo
  {author} {\bibfnamefont {A.}~\bibnamefont {Eckardt}},\ }\href {\doibase
  10.1103/PhysRevE.92.062119} {\bibfield  {journal} {\bibinfo  {journal} {Phys.
  Rev. E}\ }\textbf {\bibinfo {volume} {92}},\ \bibinfo {pages} {062119}
  (\bibinfo {year} {2015})},\ \Eprint {http://arxiv.org/abs/1508.02898}
  {arXiv:1508.02898 [cond-mat.quant-gas]} \BibitemShut {NoStop}%
\bibitem [{\citenamefont {Vorberg}\ \emph {et~al.}(2018)\citenamefont
  {Vorberg}, \citenamefont {Ketzmerick},\ and\ \citenamefont
  {Eckardt}}]{Vorberg2018}%
  \BibitemOpen
  \bibfield  {author} {\bibinfo {author} {\bibfnamefont {D.}~\bibnamefont
  {Vorberg}}, \bibinfo {author} {\bibfnamefont {R.}~\bibnamefont {Ketzmerick}},
  \ and\ \bibinfo {author} {\bibfnamefont {A.}~\bibnamefont {Eckardt}},\ }\href
  {\doibase 10.1103/PhysRevA.97.063621} {\bibfield  {journal} {\bibinfo
  {journal} {Phys. Rev. A}\ }\textbf {\bibinfo {volume} {97}},\ \bibinfo
  {pages} {063621} (\bibinfo {year} {2018})},\ \Eprint
  {http://arxiv.org/abs/1803.08866} {arXiv:1803.08866 [cond-mat.stat-mech]}
  \BibitemShut {NoStop}%
\bibitem [{foo({\natexlab{b}})}]{footnote3}%
  \BibitemOpen
  \href@noop {} {} \bibinfo {note} {In principle, the
  phenomenon of Bose selection \cite{Vorberg2013,Vorberg2015,Vorberg2018} may
  arise. If this is the case, then a single order parameter would not be
  sufficient. We do not believe that this is relevant to our problem however.
  Indeed, we describe the onset of Bose condensation, while Bose selection
  happens deep in the symmetry broken phase. In particular, our results are
  also valid in the symmetric phase. Moreover, the infinitely rapidly driven
  limit relates to the undriven limit, in that time-translation symmetry
  invariance is effectively restored, and where there is usual single mode
  condensation. Finally, although \cite{Vorberg2018} reproduces an experiment
  with a small interaction, Refs.~\cite{Vorberg2013,Vorberg2015} describe an
  ideal Bose gas. Here we describe the strongly correlated critical regime
  where interactions play a crucial role.}\BibitemShut {Stop}%
\bibitem [{\citenamefont {{Carusotto}}\ and\ \citenamefont
  {{Ciuti}}(2013)}]{Carusotto2013a}%
  \BibitemOpen
  \bibfield  {author} {\bibinfo {author} {\bibfnamefont {I.}~\bibnamefont
  {{Carusotto}}}\ and\ \bibinfo {author} {\bibfnamefont {C.}~\bibnamefont
  {{Ciuti}}},\ }\href {\doibase 10.1103/RevModPhys.85.299} {\bibfield
  {journal} {\bibinfo  {journal} {Rev. Mod. Phys.}\ }\textbf {\bibinfo {volume}
  {85}},\ \bibinfo {pages} {299} (\bibinfo {year} {2013})},\ \Eprint
  {http://arxiv.org/abs/1205.6500} {arXiv:1205.6500 [cond-mat.quant-gas]}
  \BibitemShut {NoStop}%
\bibitem [{\citenamefont {Sieberer}\ \emph {et~al.}(2016)\citenamefont
  {Sieberer}, \citenamefont {Buchhold},\ and\ \citenamefont
  {Diehl}}]{Sieberer2015b}%
  \BibitemOpen
  \bibfield  {author} {\bibinfo {author} {\bibfnamefont {L.~M.}\ \bibnamefont
  {Sieberer}}, \bibinfo {author} {\bibfnamefont {M.}~\bibnamefont {Buchhold}},
  \ and\ \bibinfo {author} {\bibfnamefont {S.}~\bibnamefont {Diehl}},\ }\href
  {http://stacks.iop.org/0034-4885/79/i=9/a=096001} {\bibfield  {journal}
  {\bibinfo  {journal} {Rep. Prog. Phys.}\ }\textbf {\bibinfo {volume} {79}},\
  \bibinfo {pages} {096001} (\bibinfo {year} {2016})},\ \Eprint
  {http://arxiv.org/abs/1512.00637} {arXiv:1512.00637 [cond-mat.quant-gas]}
  \BibitemShut {NoStop}%
\bibitem [{foo({\natexlab{c}})}]{footnote5}%
  \BibitemOpen
  \href@noop {} {} \bibinfo {note} {We have checked (cf.
  \Sect{app_x}) that the inclusion of a time-dependence in these variables does
  not change our end result.}\BibitemShut {Stop}%
\bibitem [{\citenamefont {Daley}(2014)}]{Daley2014}%
  \BibitemOpen
  \bibfield  {author} {\bibinfo {author} {\bibfnamefont {A.~J.}\ \bibnamefont
  {Daley}},\ }\href {\doibase 10.1080/00018732.2014.933502} {\bibfield
  {journal} {\bibinfo  {journal} {Adv. Phys.}\ }\textbf {\bibinfo {volume}
  {63}},\ \bibinfo {pages} {77} (\bibinfo {year} {2014})},\ \Eprint
  {http://arxiv.org/abs/1405.6694} {arXiv:1405.6694 [quant-ph]} \BibitemShut
  {NoStop}%
\bibitem [{\citenamefont {T\"auber}\ and\ \citenamefont
  {Diehl}(2014)}]{Tauber2013a}%
  \BibitemOpen
  \bibfield  {author} {\bibinfo {author} {\bibfnamefont {U.~C.}\ \bibnamefont
  {T\"auber}}\ and\ \bibinfo {author} {\bibfnamefont {S.}~\bibnamefont
  {Diehl}},\ }\href {\doibase 10.1103/PhysRevX.4.021010} {\bibfield  {journal}
  {\bibinfo  {journal} {Phys. Rev. X}\ }\textbf {\bibinfo {volume} {4}},\
  \bibinfo {pages} {021010} (\bibinfo {year} {2014})},\ \Eprint
  {http://arxiv.org/abs/1312.5182} {arXiv:1312.5182 [cond-mat.stat-mech]}
  \BibitemShut {NoStop}%
\bibitem [{\citenamefont {{Sieberer}}\ \emph {et~al.}(2013)\citenamefont
  {{Sieberer}}, \citenamefont {{Huber}}, \citenamefont {{Altman}},\ and\
  \citenamefont {{Diehl}}}]{Sieberer2013a}%
  \BibitemOpen
  \bibfield  {author} {\bibinfo {author} {\bibfnamefont {L.~M.}\ \bibnamefont
  {{Sieberer}}}, \bibinfo {author} {\bibfnamefont {S.~D.}\ \bibnamefont
  {{Huber}}}, \bibinfo {author} {\bibfnamefont {E.}~\bibnamefont {{Altman}}}, \
  and\ \bibinfo {author} {\bibfnamefont {S.}~\bibnamefont {{Diehl}}},\ }\href
  {\doibase 10.1103/PhysRevLett.110.195301} {\bibfield  {journal} {\bibinfo
  {journal} {Phys. Rev. Lett.}\ }\textbf {\bibinfo {volume} {110}},\ \bibinfo
  {eid} {195301} (\bibinfo {year} {2013})},\ \Eprint
  {http://arxiv.org/abs/1301.5854} {arXiv:1301.5854 [cond-mat.quant-gas]}
  \BibitemShut {NoStop}%
\bibitem [{\citenamefont {Kamenev}(2011)}]{kamenev2011field}%
  \BibitemOpen
  \bibfield  {author} {\bibinfo {author} {\bibfnamefont {A.}~\bibnamefont
  {Kamenev}},\ }\href@noop {} {\emph {\bibinfo {title} {Field Theory of
  Non-Equilibrium Systems}}}\ (\bibinfo  {publisher} {Cambridge University
  Press},\ \bibinfo {year} {2011})\BibitemShut {NoStop}%
\bibitem [{\citenamefont {T{\"a}uber}(2014)}]{tauber2014critical}%
  \BibitemOpen
  \bibfield  {author} {\bibinfo {author} {\bibfnamefont {U.}~\bibnamefont
  {T{\"a}uber}},\ }\href@noop {} {\emph {\bibinfo {title} {Critical Dynamics: A
  Field Theory Approach to Equilibrium and Non-Equilibrium Scaling Behavior}}}\
  (\bibinfo  {publisher} {Cambridge University Press},\ \bibinfo {year}
  {2014})\BibitemShut {NoStop}%
\bibitem [{\citenamefont {{Sieberer}}\ \emph {et~al.}(2015)\citenamefont
  {{Sieberer}}, \citenamefont {{Chiocchetta}}, \citenamefont {{Gambassi}},
  \citenamefont {{T{\"a}uber}},\ and\ \citenamefont {{Diehl}}}]{Sieberer2015}%
  \BibitemOpen
  \bibfield  {author} {\bibinfo {author} {\bibfnamefont {L.~M.}\ \bibnamefont
  {{Sieberer}}}, \bibinfo {author} {\bibfnamefont {A.}~\bibnamefont
  {{Chiocchetta}}}, \bibinfo {author} {\bibfnamefont {A.}~\bibnamefont
  {{Gambassi}}}, \bibinfo {author} {\bibfnamefont {U.~C.}\ \bibnamefont
  {{T{\"a}uber}}}, \ and\ \bibinfo {author} {\bibfnamefont {S.}~\bibnamefont
  {{Diehl}}},\ }\href {\doibase 10.1103/PhysRevB.92.134307} {\bibfield
  {journal} {\bibinfo  {journal} {Phys. Rev. B}\ }\textbf {\bibinfo {volume}
  {92}},\ \bibinfo {eid} {134307} (\bibinfo {year} {2015})},\ \Eprint
  {http://arxiv.org/abs/1505.00912} {arXiv:1505.00912 [cond-mat.stat-mech]}
  \BibitemShut {NoStop}%
\bibitem [{\citenamefont {Aron}\ \emph {et~al.}(2018)\citenamefont {Aron},
  \citenamefont {Biroli},\ and\ \citenamefont {Cugliandolo}}]{Aron2018}%
  \BibitemOpen
  \bibfield  {author} {\bibinfo {author} {\bibfnamefont {C.}~\bibnamefont
  {Aron}}, \bibinfo {author} {\bibfnamefont {G.}~\bibnamefont {Biroli}}, \ and\
  \bibinfo {author} {\bibfnamefont {L.~F.}\ \bibnamefont {Cugliandolo}},\
  }\href {\doibase 10.21468/SciPostPhys.4.1.008} {\bibfield  {journal}
  {\bibinfo  {journal} {SciPost Phys.}\ }\textbf {\bibinfo {volume} {4}},\
  \bibinfo {pages} {008} (\bibinfo {year} {2018})},\ \Eprint
  {http://arxiv.org/abs/1705.10800} {arXiv:1705.10800 [cond-mat.stat-mech]}
  \BibitemShut {NoStop}%
\bibitem [{\citenamefont {Arrachea}(2005)}]{Arrachea2005}%
  \BibitemOpen
  \bibfield  {author} {\bibinfo {author} {\bibfnamefont {L.}~\bibnamefont
  {Arrachea}},\ }\href {\doibase 10.1103/PhysRevB.72.125349} {\bibfield
  {journal} {\bibinfo  {journal} {Phys. Rev. B}\ }\textbf {\bibinfo {volume}
  {72}},\ \bibinfo {pages} {125349} (\bibinfo {year} {2005})},\ \Eprint
  {http://arxiv.org/abs/cond-mat/0505153} {cond-mat/0505153} \BibitemShut
  {NoStop}%
\bibitem [{\citenamefont {Wu}\ and\ \citenamefont {Cao}(2008)}]{wu2008}%
  \BibitemOpen
  \bibfield  {author} {\bibinfo {author} {\bibfnamefont {B.~H.}\ \bibnamefont
  {Wu}}\ and\ \bibinfo {author} {\bibfnamefont {J.~C.}\ \bibnamefont {Cao}},\
  }\href {\doibase 10.1088/0953-8984/20/8/085224} {\bibfield  {journal}
  {\bibinfo  {journal} {J. Phys. Condens. Matter}\ }\textbf {\bibinfo {volume}
  {20}},\ \bibinfo {pages} {085224} (\bibinfo {year} {2008})}\BibitemShut
  {NoStop}%
\bibitem [{\citenamefont {Stefanucci}\ \emph {et~al.}(2008)\citenamefont
  {Stefanucci}, \citenamefont {Kurth}, \citenamefont {Rubio},\ and\
  \citenamefont {Gross}}]{Stefanucci2008}%
  \BibitemOpen
  \bibfield  {author} {\bibinfo {author} {\bibfnamefont {G.}~\bibnamefont
  {Stefanucci}}, \bibinfo {author} {\bibfnamefont {S.}~\bibnamefont {Kurth}},
  \bibinfo {author} {\bibfnamefont {A.}~\bibnamefont {Rubio}}, \ and\ \bibinfo
  {author} {\bibfnamefont {E.~K.~U.}\ \bibnamefont {Gross}},\ }\href {\doibase
  10.1103/PhysRevB.77.075339} {\bibfield  {journal} {\bibinfo  {journal} {Phys.
  Rev. B}\ }\textbf {\bibinfo {volume} {77}},\ \bibinfo {pages} {075339}
  (\bibinfo {year} {2008})},\ \Eprint {http://arxiv.org/abs/cond-mat/0701279}
  {cond-mat/0701279} \BibitemShut {NoStop}%
\bibitem [{\citenamefont {Tsuji}\ \emph {et~al.}(2008)\citenamefont {Tsuji},
  \citenamefont {Oka},\ and\ \citenamefont {Aoki}}]{Tsuji2008}%
  \BibitemOpen
  \bibfield  {author} {\bibinfo {author} {\bibfnamefont {N.}~\bibnamefont
  {Tsuji}}, \bibinfo {author} {\bibfnamefont {T.}~\bibnamefont {Oka}}, \ and\
  \bibinfo {author} {\bibfnamefont {H.}~\bibnamefont {Aoki}},\ }\href {\doibase
  10.1103/PhysRevB.78.235124} {\bibfield  {journal} {\bibinfo  {journal} {Phys.
  Rev. B}\ }\textbf {\bibinfo {volume} {78}},\ \bibinfo {pages} {235124}
  (\bibinfo {year} {2008})},\ \Eprint {http://arxiv.org/abs/0808.0379}
  {arXiv:0808.0379 [cond-mat.str-el]} \BibitemShut {NoStop}%
\bibitem [{foo({\natexlab{d}})}]{footnote}%
  \BibitemOpen
  \href@noop {} {} \bibinfo {note} {The phase with broken
  $U(1)$ symmetry however exhibits gapless Goldstone modes. In $d>2$
  dimensions, the associated IR fluctuations are phase space suppressed and do
  not destroy the ordered phase of an undriven system. Because the divergences
  emerging from the drive are copies of the single undriven divergence, we do
  not expect the presence of a drive to alter this behavior
  qualitatively.}\BibitemShut {Stop}%
\bibitem [{\citenamefont {Kibble}(1976)}]{kibble1976}%
  \BibitemOpen
  \bibfield  {author} {\bibinfo {author} {\bibfnamefont {T.~W.~B.}\
  \bibnamefont {Kibble}},\ }\href {\doibase 10.1088/0305-4470/9/8/029}
  {\bibfield  {journal} {\bibinfo  {journal} {J. Phys. A: Math. Gen.}\ }\textbf
  {\bibinfo {volume} {9}},\ \bibinfo {pages} {1387} (\bibinfo {year}
  {1976})}\BibitemShut {NoStop}%
\bibitem [{\citenamefont {Zurek}(1985)}]{Zurek1985}%
  \BibitemOpen
  \bibfield  {author} {\bibinfo {author} {\bibfnamefont {W.~H.}\ \bibnamefont
  {Zurek}},\ }\href {\doibase 10.1038/317505a0} {\bibfield  {journal} {\bibinfo
   {journal} {Nature}\ }\textbf {\bibinfo {volume} {317}},\ \bibinfo {pages}
  {505} (\bibinfo {year} {1985})}\BibitemShut {NoStop}%
\bibitem [{\citenamefont {Sieberer}\ \emph {et~al.}(2014)\citenamefont
  {Sieberer}, \citenamefont {Huber}, \citenamefont {Altman},\ and\
  \citenamefont {Diehl}}]{Sieberer2013b}%
  \BibitemOpen
  \bibfield  {author} {\bibinfo {author} {\bibfnamefont {L.~M.}\ \bibnamefont
  {Sieberer}}, \bibinfo {author} {\bibfnamefont {S.~D.}\ \bibnamefont {Huber}},
  \bibinfo {author} {\bibfnamefont {E.}~\bibnamefont {Altman}}, \ and\ \bibinfo
  {author} {\bibfnamefont {S.}~\bibnamefont {Diehl}},\ }\href {\doibase
  10.1103/PhysRevB.89.134310} {\bibfield  {journal} {\bibinfo  {journal} {Phys.
  Rev. B}\ }\textbf {\bibinfo {volume} {89}},\ \bibinfo {pages} {134310}
  (\bibinfo {year} {2014})},\ \Eprint {http://arxiv.org/abs/1309.7027}
  {arXiv:1309.7027 [cond-mat.quant-gas]} \BibitemShut {NoStop}%
\end{thebibliography}
\end{document}